\documentclass[letterpaper, 11pt, leqno]{article}
\usepackage{paper}
\usepackage{math}
\hypersetup{pdftitle={Learning to Detect Recessions from Pseudo-out-of-sample Tests}}
\newcommand{\bib}{paper.bib}
\newcommand{\pdf}{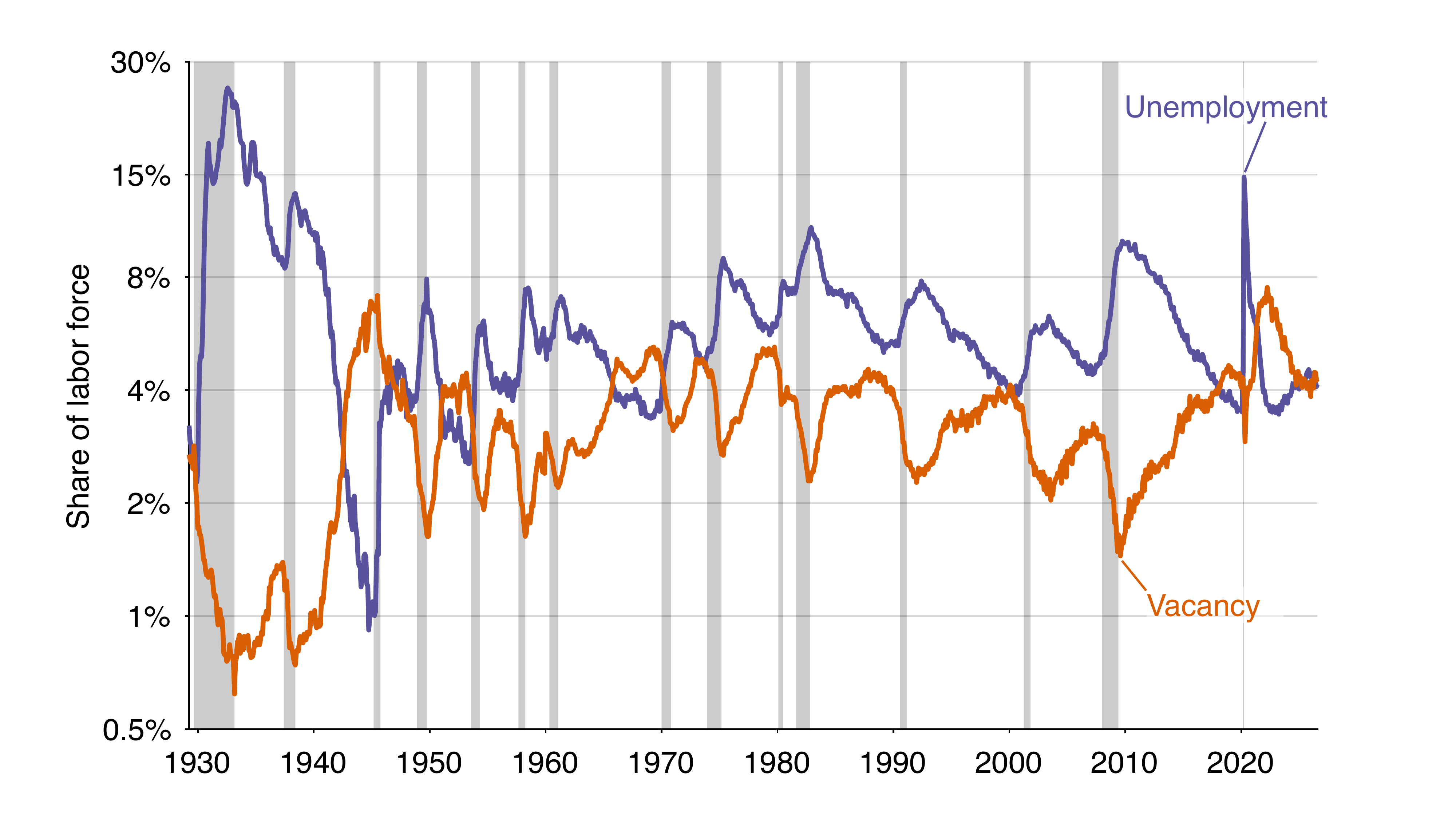}
\begin{document}

\title{Learning to Detect Recessions\\from Pseudo-out-of-sample Tests}
\author{Pascal Michaillat
\thanks{University of California, Santa Cruz. I thank Clement Bohr, Yuriy Gorodnichenko, Manfred Keil, Michael Leung, Adam McCloskey, Jeremy Piger, Mahdi Rastad, Emmanuel Saez, Dick Startz, and Eduardo Zambrano for helpful comments.}}
\date{October 2026} 
\paperurl{https://pascalmichaillat.org/17/}
\begin{titlepage}
\maketitle

Everyone is keen to know when a recession has started, but official announcements come only after several months, and real-time detection methods typically struggle---as the Great Recession showed. These methods face two challenges. First, they rely on data that are later significantly revised (such as GDP), so they only have a muddied view of the situation in real time. Second, they optimize in-sample performance, but this does not guarantee out-of-sample results. This paper proposes an algorithm that addresses both. First, it detects recessions from data on labor market slack---unemployment and vacancy rates---which are barely revised. Second, it selects recession classifiers for out-of-sample performance, using pseudo-out-of-sample tests on past recessions. Placed in the shoes of the NBER Business Cycle Dating Committee, the algorithm issues an early warning 5.2 months before the NBER announcement on average over the 6 recessions since 1978, and it reaches a 100\% recession probability before the announcement or in the same month in 5 of them. The algorithm produces only one false positive out of sample: it takes the 2023--2024 labor market slowdown for a recession. Trained on the data between April 1929 and December 2025, the algorithm gives a recession probability of 0\% in August 2026.

\end{titlepage}

\section{Introduction}

\paragraph{Recessions are the harbinger of doom} When the economy enters a recession, economic conditions are about to deteriorate significantly and rapidly. Because the damage is done quickly and repaired slowly, households, firms, and policymakers all want to know as soon as possible whether a recession has started. Hence the attention that recessions receive. As \citet[p.~8]{RR20} note: 
\begin{quote}
Recessions also play an important role in economic discussions among policymakers, the public, financial professionals, and the media. For example, the Washington Post published 630 items that featured the term ``recession'' in 2019, the New York Times 818, and the Wall Street Journal 1037.
\end{quote}

\paragraph{Official recession announcements take time} In the United States, recessions are officially identified by the Business Cycle Dating Committee of the National Bureau of Economic Research \citep{NBER23}. But the committee announces recessions with a substantial delay: 6.3 months after the recession's onset on average, and as late as 11 months for the Great Recession (table~\ref{t:nber}). As the \citet{NBER08a} explains, the delay is deliberate:
\begin{quote}
The committee's approach to determining the dates of turning points is retrospective. We wait until sufficient data are available to avoid the need for major revisions. In particular, in determining the date of a peak in activity, and thus the onset of recession, we wait until we are confident that, even in the event that activity
begins to rise again immediately, it has declined enough to meet the criterion of depth. As a result, we tend to wait to identify a peak until many months after it actually occurs.
\end{quote}
Because its dates become the official chronology of US business cycles, the committee must be sure before it speaks. Therefore, the announcements only come once the economy has cooled markedly. Because of this delay, waiting for the NBER to announce a recession before acting is often not practical for policymakers---or for businesses and households.

\paragraph{Challenges of detecting the Great Recession in real time} Because the NBER is slow, a vast academic literature, and a substantial private-sector effort, is devoted to detecting recessions in real time.\footnote{See for instance \citet{H89}, \citet{SW89}, \citet{C98a}, \citet{CH06}, \citet{CP08}, \citet{L08}, \citet{GP17}, \citet{HS20}, and \citet{FG24}. Other algorithms attempt to predict future US recessions at various horizons: for instance, \citet{SW93}, \citet{EM98}, \citet{Q01}, \citet{CP05}, \citet{D05}, \citet{KS08a}, \citet{N10}, \citet{CIL11}, \citet{N14}, \citet{CEM14}, \citet{F16}, \citet{VGV21}, \citet{L24}, and \citet{PSW26}.} Yet, as \citet{H11b} documents, these methods did not do well when the Great Recession came.\footnote{The last recession posed no such challenge. At the onset of the pandemic, between February and April 2020, the unemployment rate rose from 3.5\% to 14.8\% \citep{UNRATE}, labor market tightness collapsed from 1.24 to 0.26 \citep{ABCD}, and the industrial production index fell from 101.4 to 84.6 \citep{INDPRO}. There was no doubt that a recession had started.} The recession probability given by the yield-curve model of \citet{W06} peaked at 50\% in April 2007 and then declined steadily; it stood at 0.1\% on August 20, 2008, ``just before one of the worst 6 month periods that the economy has experienced over the last half-century'' \citep[p.~1013]{H11b}. In August 2008, the threshold rules of \citet{L08}, based on employment, industrial production, and unemployment, indicated that the data through June did not yet signal a recession---although the recession had already started in January \citep[pp.~1024--1025]{H11b}. The Markov-switching model of four monthly indicators developed by \citet{CP08} and the Markov-switching model of GDP developed by \citet{CH06} signaled the recession in January 2009---after the announcement by the NBER in December 2008. The GDP-based declaration came ``much later than all observers understood that a powerful economic downturn was underway'' \citep[p.~1017]{H11b}. The performance of recession detection algorithms during the Great Recession indicates that more work is required to be ready for the next recession.

\begin{table}[t]
\caption{\citet{NBER21} Business Cycle Dating Committee recession announcements}
\begin{tabular*}{\textwidth}{@{\extracolsep{\fill}}*{3}{c}}
\toprule
Recession start & Announcement date & Announcement delay (months) \\
\midrule
February 1980  & June 1980     & 4  \\
August 1981     & January 1982  & 5   \\
August 1990     & April 1991    & 8   \\
April 2001    & November 2001 & 7  \\
January 2008 & December 2008 & 11 \\
March 2020 & June 2020     & 3   \\
\midrule
& Mean delay:& 6.3\\
\bottomrule
\end{tabular*}
\label{t:nber}\end{table}

\paragraph{General challenges of detecting recessions in real time} Overall, as \citet[p.~1006]{H11b} explains, detecting recessions ``on a real time, out-of-sample basis is a bigger challenge than many academics might assume.'' The first challenge is that ``the data available at the time often send different signals to what one sees once the data are subsequently revised'' \citep[p.~1009]{H11b}. For instance, as reported in January 2002, the GDP numbers sent no clear recession signal; but as revised by the end of 2002, real GDP had declined in three successive quarters of 2001, ``which certainly sounds unambiguously like a recession.'' The second challenge is that a classifier that worked in the past may not work in the future. After reviewing the real-time track record of two prominent forecasting models, \citet[p.~1013]{H11b} concludes that ``a good in-sample fit is no guarantee of a good out-of-sample real-time performance.''  This paper aims to address these two challenges.

\paragraph{Finding data reliable in real time: labor market slack} Existing detection methods principally use data on quantities---such as GDP, sales, employment, and industrial production. This paper instead uses data on slack. As fluctuations in quantities are largely explained by fluctuations in slack, fluctuations in slack are salient markers of recessions \citep[chapter~2]{M26}. As this is the form of slack that is measured best, the paper detects recessions using labor market slack: specifically, using the unemployment rate and job vacancy rate. The unemployment and vacancy rates mark recessions clearly: at the onset of any recession, unemployment rises rapidly while vacancies plummet (figure~\ref{f:data}). The unemployment and vacancy rates have several advantages for this paper's purpose:
\begin{enumerate}
\item They are barely revised. The vacancy rate is revised one month after its initial release, but after that revisions are minimal, so the information provided in real time is ``almost indistinguishable'' from the information provided in the final version \citep{CGL20a}. This is unsurprising because slack is easier to measure in real time than quantities: counting jobseekers and job openings is much easier than computing all the value added in the economy.
\item They are available every month since 1929, covering 15 recessions \citep{PZ21}. The two series therefore provide ample data to train the recession detection algorithm and conduct pseudo-out-of-sample tests. 
\item They are measured independently. Unemployment comes from a household survey: the Current Population Survey (CPS) conducted by the \citet{UNEMPLOY}. Vacancies come from help-wanted advertising indices in the past and from a firm survey today: the Job Openings and Labor Turnover Survey (JOLTS) conducted by the \citet{JTSJOL}. Combining them therefore yields a recession indicator that is less noisy than unemployment alone \citep{MS25}.
\item They do not grow along a trend, so they never need to be detrended, and they therefore provide a cleaner, more transparent recessionary signal, especially in real time (see \citealt{H18} for issues that typically arise with detrending).
\item Theory tells us that over typical business cycles, supply-side slack (unemployment) and demand-side slack (vacancies) move in opposite directions: in recessions, unemployment rises and vacancies fall as the economy moves along the Beveridge curve \citep{M26}.
\end{enumerate} 

\paragraph{Improving out-of-sample performance: pseudo-out-of-sample tests} Even with labor market slack data, there are many ways to build recession classifiers. How should the data be filtered? Which thresholds should trigger recession warnings? A number of threshold rules based on the unemployment rate, such as the \citet{S19} rule, have been developed and are quite effective \citep{CGL20b}.\footnote{Other unemployment-based threshold rules have been developed by \citet{S08}, \citet{HS12a}, \citet{SFH21}, \citet{P24}, \citet{OS25}, \citet{S25}, and \citet{GJS26}. Such rules work well in other countries too \citep{MB26}.} Recession classifiers are typically selected for their in-sample performance: \citet[p.~1047]{MS25} argue that their Michez rule improves on the Sahm rule in part because ``this new rule detects recessions faster than the Sahm rule: with an average delay of 1.2 months instead of 2.7 months, and a maximum delay of 3 months instead of 7 months.'' The goal of recession detection is to perform well out of sample, however, and recession classifiers should be selected for that goal. Building on \citet{SW03a,SW09a}, I therefore select recession classifiers based on their performance in a series of pseudo-out-of-sample tests, in the hope of building a classifier that will do well when the next recession comes.\footnote{See also \citet{CM13} and \citet{D15} for discussions of pseudo-out-of-sample testing.}

\paragraph{Recession detection algorithm}  The algorithm proceeds in three steps. First, it constructs 95,832 recession indicators by filtering and combining unemployment and vacancy data in many different ways. The aim is to extract the best possible recessionary signal from the data.\footnote{This algorithm aims to learn more by filtering two data series in a larger number of ways. By contrast, the following papers attempt to learn more by processing a larger set of data: \citet{CIL11} consider 141 time series; \citet{N14} considers 1,500 series (132 distinct series and their lags); \citet{F16} considers 133 series constructed by \citet{MN16}; \citet{PSW26} complement hard-data series with 4 soft-data series reflecting the sentiments of journalists, policymakers, businesses, and households; and \citet{S25} use 55 series including measures of unemployment disaggregated by race, education, age, sector, and location. This paper's focus on a small number of predictive variables fits with \citet{P20}'s assessment that a narrow set of variables classify US business cycles well.} Second, inspired by \citet{MS25}, it builds probabilistic recession classifiers that rely on two detection thresholds: the recession probability turns strictly positive when the indicator crosses the lower threshold (early warning) and reaches 1 when it crosses the upper threshold (final warning). The threshold band captures the uncertainty about where the boundary between recessions and nonrecessions lies. The algorithm only retains perfect classifiers: those with no false negatives (existent but undetected recessions) and no false positives (nonexistent but detected recessions). Despite the seeming flexibility of the filtering, perfection is hard to achieve: only 1,028 of the 95,832 indicators yield a perfect classifier over 1929--2025. But 1,028 classifiers is still many, and every one of them fits the training data equally well in that it makes no in-sample errors. The third step selects among the perfect classifiers. Since they all fit the data equally well, the algorithm asks which way of choosing among them would have worked on episodes it had not seen. It splits the past data at several dates, lets each of 34 selection methods---based on qualities such as the earliness, robustness, and accuracy of the classifiers---choose classifiers from the data before the split, and scores those classifiers on the episode after it. The method with the best average score in these pseudo-out-of-sample tests is applied to all the data, and the classifier that it selects detects the next recession.

\paragraph{Placing the algorithm in the shoes of the NBER} To assess how well this procedure would have worked in practice, I put the algorithm in the shoes of the NBER Business Cycle Dating Committee. I simulate an algorithm that faces, one after the other, the 6 episodes that the committee has faced since its creation in 1978: the twin Volcker recessions, the Gulf War recession, the dot-com recession, the Great Recession, the pandemic recession, and the 2023--2024 labor market slowdown, which was not a recession but which many detection rules took for one. This simulation is the out-of-sample evaluation of the algorithm, and the pseudo-out-of-sample tests sit inside it. Each out-of-sample test uses only the data available at its start, both to find the perfect classifiers and to run the pseudo-out-of-sample tests that rank the selection methods. Once it is over, its window becomes one more pseudo-out-of-sample test for the later out-of-sample tests. The exercise is out of sample in this sense only: of course I had seen the full history of the data before starting to work on the paper.

\paragraph{Example of pseudo-out-of-sample test} Consider the algorithm in January 1978, when the NBER committee was created, so the algorithm stands where the committee stood. The algorithm has data from April 1929 to December 1977. To choose a selection method, it runs a pseudo-out-of-sample test on the most recent recessionary period, January 1973--December 1977. It finds the perfect classifiers on April 1929--December 1972 and evaluates them on several qualities: how early their early and final warnings come, how widely they separate recessions from nonrecessions, and so on. Each of the 34 selection methods picks a classifier or an ensemble of classifiers based on these qualities, and each pick is scored on the 1973--1975 recession. The method that picks the classifier with the widest margin between recessions and nonrecessions does best. The algorithm therefore applies that method to April 1929--December 1977, and from January 1978 it uses the selected classifier to detect recessions out of sample. In January 1985, once the twin Volcker recessions are over, January 1978--December 1984 becomes a second pseudo-out-of-sample test. The algorithm re-ranks the methods by their average score on the two pseudo-out-of-sample tests and switches to the selection method that is now best.

\paragraph{Out-of-sample performance of the algorithm compared to the NBER} The performance of the algorithm is strikingly good. Although the algorithm is trained only on past data, it issues an early warning before the NBER announcement in each of the 6 recessions that it faces, by 5.2 months on average. But the NBER Business Cycle Dating Committee does well, too: the algorithm's final warnings tend to come around the same time as the NBER announcements---4.8 months after the recession onset on average, against 6.3 months for the announcements. Being certain that a recession has started takes time, for the algorithm as for the committee. The choices made by the algorithm are surprisingly stable: the algorithm keeps the same classifier from 1995 to 2022; that classifier detects the dot-com recession, Great Recession, and pandemic recession before the NBER. 

\paragraph{Current state of the algorithm} Averaged over all pseudo-out-of-sample tests, the best selection method selects the single classifier with the earliest final warnings. Methods based on the earliness of early warnings are the worst---which shows that classifiers selected to provide rapid early warnings in sample do not necessarily perform well out of sample. The 2023--2024 labor market slowdown had a notable impact: the classifier used from 1995 to 2022 took the slowdown for a recession, and it is no longer perfect once the slowdown is in the training data. Trained on April 1929--December 2025 data, the algorithm now selects a minimum indicator of level changes in unemployment and vacancy rates, as in the Michez rule. Since 1978, its recession probability starts rising 2.2 months after recession onset and reaches 1 after 2.8 months on average, so its final warnings precede the NBER announcements by 3.5 months. 

\paragraph{Application of the algorithm in real time} Going forward, the algorithm is simple to apply. Each month, once the BLS releases the new unemployment and vacancy data, the indicator is updated and compared to the two thresholds. When the indicator crosses the lower threshold, the recession probability turns positive; between the thresholds, the probability is between 0 and 1; when it crosses the upper threshold, the recession probability becomes 1. For now, the selected classifier gives a recession probability of 0\% in August 2026.

\section{Data}

This section presents the data on US recessions and on the two measures of labor market slack used in the paper: the unemployment rate and the vacancy rate. The data cover April 1929 to August 2026, which is the longest period for which the data are available \citep[p.~350]{MS24}.

\subsection{Recession dates}

US recessions are officially identified by the Business Cycle Dating Committee of the NBER, which was established in 1978 \citep[p.~14]{RR20}. The \citet{NBER24} identifies the peaks and troughs of US business cycles by looking holistically at numerous macroeconomic variables. By convention, the first month of a recession is the month following the peak and the last month of a recession is the month of the trough. Between April 1929 and August 2026, the \citet{NBER23} identifies 15 recessions. The NBER-dated recessions are displayed in figure~\ref{f:data}.

\subsection{Unemployment rate} 

Between April 1929 and December 1947, I use the monthly unemployment rate constructed by \citet{PZ21}. They extrapolate \citet{W92a}'s annual unemployment series to a monthly series using monthly unemployment rates compiled by the NBER. Between January 1948 and August 2026, the unemployment rate is computed in the usual way: it is the number of jobseekers measured by the \citet{UNEMPLOY} from the CPS, divided by the civilian labor force measured by the \citet{CLF16OV} from the CPS. This is the standard, official measure of unemployment, labeled U3 by the \citet{BLS23}. The unemployment rate used in the analysis is plotted in figure~\ref{f:data}. It is countercyclical, rising sharply at the onset of all recessions.

\begin{figure}[t]
\widegraphic[1]{\pdf}
\caption{Monthly US unemployment and vacancy rates, April 1929--August 2026}
\note{The unemployment rate is computed from data produced by \citet{PZ21} and the \citet{UNEMPLOY,CLF16OV}. The vacancy rate is computed from data produced by \citet{PZ21}, \citet{B10}, and the \citet{CLF16OV,JTSJOL}. Shaded areas indicate recessions dated by the \citet{NBER23}.}
\label{f:data}\end{figure}

\subsection{Vacancy rate}\label{s:vacancy}

Between April 1929 and December 1950, the vacancy rate is based on the help-wanted index created by the Metropolitan Life Insurance Company. This index aggregates help-wanted advertisements from newspapers across major US cities. It is considered a reliable proxy for job vacancies \citep{Z98a}. The index is scaled to align with \citet{B10}'s vacancy rate at the end of 1950, effectively translating the index into a vacancy rate.

Between January 1951 and December 2000, I use the vacancy rate produced by \citet{B10}. This series is based on the Conference Board's help-wanted advertising index, adjusted to account for the shift from print advertising to online advertising in the 1990s. The Conference Board index aggregates help-wanted advertising in major metropolitan newspapers in the United States. It serves as a reliable proxy for job vacancies \citep{A87,S05}. The Conference Board index is scaled to align with the JOLTS vacancy rate in 2001, effectively translating the index into a vacancy rate.

Between January 2001 and August 2026, I compute the vacancy rate as the number of job openings measured by the \citet{JTSJOL} from the JOLTS, divided by the size of the labor force measured by the \citet{CLF16OV} from the CPS. To best align labor force and vacancy data, following \citet[p.~346]{MS24}, I shift forward by one month the number of job openings from JOLTS. For instance, I assign to April 2025 the number of job openings that the BLS assigns to March 2025.

I then splice the three vacancy series to create a continuous vacancy rate covering April 1929--August 2026. The vacancy rate is plotted in figure~\ref{f:data}. It is procyclical, dropping sharply at the onset of all recessions.

\subsection{Availability and revisions of labor market data}\label{s:revisions}

The unemployment and vacancy data required to apply the algorithm in any given month are released in the first week of the following month, usually on a Tuesday for the JOLTS data and on a Friday for the CPS data \citep{BLS24b}. Because job openings are shifted forward by one month (section~\ref{s:vacancy}), the JOLTS release used for month $t$ is the one covering month $t-1$, which the BLS publishes in the first week of month $t+1$. Thus, the algorithm can be applied in real time.

The recession probability constructed in real time might not be its final value because the unemployment and vacancy data are revised after their initial release. The number of job openings released by the \citet{JTSJOL} is preliminary and updated one month after its initial release, to incorporate additional survey responses received from businesses and government agencies \citep{BLS24c}. Additionally, the BLS revises the prior 5 years of CPS and JOLTS data each year at the beginning of January, to account for revisions to seasonal factors, population estimates, and employment estimates, but these revisions are typically minor \citep{BLS24c,BLS25a}.
	
\section{Construction of the recession indicators}\label{s:construction}

This section constructs 95,832 recession indicators by filtering and combining the unemployment and vacancy data. My motivation is that unemployment and vacancy data capture recessions almost perfectly (as is apparent in figure~\ref{f:data}), so instead of looking at other data sources, I try to make progress by determining the best possible lens through which to look at the data. The indicators are created in 4 steps: smoothing the data, detecting turning points, scaling variations, and combining single-variable indicators.

\subsection{Smoothing the data}

I smooth data by moving average. I first compute simple trailing averages of the unemployment rate:
\begin{equation}
\bar{u}(t) = \frac{\sum_{k=0}^{\a} u(t-k)}{\a+1},
\label{e:usma}\end{equation}
where $\a = 0, 1, 2, \ldots, 11$ determines the amount of smoothing. The case $\a = 0$ corresponds to no smoothing; the case $\a = 11$ corresponds to a 12-month trailing average. The Sahm and Michez rules set $\a = 2$, which corresponds to a 3-month trailing average.

The vacancy rate is smoothed in the same way: 
\begin{equation}
\bar{v}(t) = \frac{\sum_{k=0}^{\a} v(t-k)}{\a+1},
\label{e:vsma}\end{equation}
where $\a = 0, 1, 2, \ldots, 11$.

As an alternative, I also use an exponentially weighted moving average, which is defined recursively by
\begin{equation}
\bar{u}(t) = \a u(t) + (1-\a) \bar{u}(t-1),
\label{e:uema}\end{equation}
where $\a = 0.1, 0.2, 0.3, \ldots, 1$ determines the amount of smoothing. The case $\a = 1$ corresponds to no smoothing, and lower values of $\a$ generate more smoothing.

The vacancy rate is smoothed in the same way:
\begin{equation}
\bar{v}(t) = \a v(t) + (1-\a) \bar{v}(t-1),
\label{e:vema}\end{equation}
where $\a = 0.1, 0.2, 0.3, \ldots, 1$.

\subsection{Detecting turning points}

To detect turning points in the unemployment rate, I isolate the minimum of the unemployment rate at various horizons: 
\begin{equation}
u^{\min}(t) = \min[0\leq k \leq \b] \bar{u}(t-k),
\label{e:umin}\end{equation}
where $\b = 1, 2, 3, \ldots, 18$ months is the horizon of the trailing minimum. The Sahm and Michez rules use $\b = 12$, which corresponds to a 12-month trailing minimum.

Then, the increase in unemployment rate from the turning point is computed as
\begin{equation}
\tilde{u}(t) = \bar{u}(t) - u^{\min}(t).
\label{e:utilde}\end{equation}
Thus the current unemployment rate is compared to values ranging from last month's unemployment rate to the minimum unemployment rate over the past 18 months.

I proceed analogously to determine turning points in the vacancy rate. I first take the maximum of the vacancy rate at various monthly horizons:
\begin{equation}
v^{\max}(t) = \max[0\leq k \leq \b] \bar{v}(t-k),
\label{e:vmax}\end{equation}
where $\b = 1, 2, 3, \ldots, 18$ months. Then, the decrease in vacancy rate from the turning point is computed as
\begin{equation}
\tilde{v}(t) = v^{\max}(t) - \bar{v}(t).
\label{e:vtilde}\end{equation}
The current vacancy rate is compared to values ranging from last month's vacancy rate to the maximum vacancy rate over the past 18 months.

\subsection{Scaling variations}

The unemployment and vacancy variations $\tilde{u}(t)$ and $\tilde{v}(t)$ measure level changes in the unemployment and vacancy rates. It is not entirely clear, however, whether what matters are level changes or percentage changes. Maybe recessions occur when the unemployment rate increases by 1pp, but it is just as possible that recessions occur when the unemployment rate increases by 10\%. Since I am looking for the best recession indicator, I do not want to restrict the type of indicators that I consider. 

So in addition to level changes in unemployment and vacancy rates, I also consider percentage changes and intermediate changes. To do that, I consider Box-Cox transformations of the unemployment indicator. Formally, I consider all the indicators of the form
\begin{equation}
\hat{u}(t) = \frac{\bs{\bar{u}(t)}^{\g}-\bs{u^{\min}(t)}^{\g}}{\g} \approx \frac{\tilde{u}(t)}{u^{\min}(t)^{1-\g}},
\label{e:uboxcox}\end{equation}
where the Box-Cox parameter takes values $\g = 0, 0.1, 0.2, \ldots, 1$ (figure~\ref{f:indicators}). The case $\g=0$ reduces to percentage changes: 
\begin{equation*}
\hat{u}(t) = \log\of{\frac{\bar{u}(t)}{u^{\min}(t)}} \approx \frac{\tilde{u}(t)}{u^{\min}(t)}.
\end{equation*}
The case $\g=1$ reduces to level changes: 
\begin{equation*}
\hat{u}(t) = \tilde{u}(t).
\end{equation*}
The Sahm and Michez rules both consider level changes, so $\g=1$.

Similarly, I consider Box-Cox transformations of the vacancy indicator. That is, I construct all the vacancy indicators of the form
\begin{equation}
\hat{v}(t) = \frac{\bs{v^{\max}(t)}^{\g} - \bs{\bar{v}(t)}^{\g}}{\g}\approx \frac{\tilde{v}(t)}{v^{\max}(t)^{1-\g}},
\label{e:vboxcox}\end{equation}
where $\g = 0, 0.1, 0.2, \ldots, 1$. 

\subsection{Combining single-variable indicators}

\begin{figure}[t]
\subcaptionbox{4,356 unemployment indicators \label{f:u}}{\smallgraphic[2]{\pdf}}\hfill
\subcaptionbox{4,356 vacancy indicators \label{f:v}}{\smallgraphic[3]{\pdf}}
\caption{Unemployment and vacancy indicators obtained from raw data}
\note{The unemployment and vacancy indicators are computed as described in section~\ref{s:construction} from the unemployment and vacancy rates in figure~\ref{f:data}. Shaded areas indicate recessions dated by the \citet{NBER23}.}
\label{f:indicators}\end{figure}

The last step is to combine the unemployment and vacancy indicators constructed previously and displayed in figure~\ref{f:indicators}.

Besides using all the individual unemployment and vacancy indicators, I construct new  indicators that linearly combine the unemployment and vacancy indicators filtered in the same way:
\begin{equation}
i(t) = \d \hat{u}(t) + (1-\d) \hat{v}(t),
\label{e:uv}\end{equation}
where $\d = 0, 0.1, 0.2 \ldots, 1$ is the weight on the unemployment indicator, and the individual indicators $\hat{u}(t)$ and $\hat{v}(t)$ are obtained using the same filtering parameters $\a$, $\b$, and $\g$. The Sahm rule only uses the unemployment indicator, which corresponds to $\d = 1$.

Moreover, \citet{MS25} showed that the minimum of the unemployment and vacancy indicators performs well to detect recessions because it is a less noisy signal of recessions than either the unemployment indicator or the vacancy indicator alone. Motivated by this insight, I also consider the minimum and maximum of the unemployment and vacancy indicators, as well as their linear combinations:
\begin{equation}
i(t) = \d \min{\hat{u}(t),\hat{v}(t)} + (1-\d) \max{\hat{u}(t),\hat{v}(t)},
\label{e:minmax}\end{equation}
where $\d = 0, 0.1, 0.2, \ldots, 1$ is the weight on the minimum indicator, and here too the individual indicators $\hat{u}(t)$ and $\hat{v}(t)$ are obtained using the same filtering parameters $\a$, $\b$, and $\g$. The Michez rule only uses the minimum indicator, which corresponds to $\d = 1$.

\subsection{Summary}

Overall, from the unemployment rate, I construct $[12+10]\times 18 \times 11 = $ 4,356  recession indicators. Similarly, from the vacancy rate, I construct another 4,356 recession indicators. By combining these indicators linearly, I produce a total of 4,356 $\times 11 =$ 47,916 indicators. By combining the minimum and maximum of these indicators, I produce another batch of 47,916 indicators. So in total, I have 47,916 $+$ 47,916 $=$  95,832 recession indicators, from which I will build recession classifiers.

\section{Perfect probabilistic recession classifiers}\label{s:ensemble}

Having constructed the 95,832 recession indicators, I pair them with thresholds to form classifiers, retain the classifiers that are perfect over the training period, and turn them into recession probabilities. Throughout the section, the training period is April 1929--December 2025; section~\ref{s:simulation} reruns the entire procedure on shorter training periods.

\subsection{Detection methodology}	

A specific classifier is a specific indicator with a specific threshold. The classifier delineates periods as recessions by identifying points where the indicator crosses the threshold from below, contingent upon the economy being in expansion prior to the crossing. The classifier maintains a state variable for expansion and recession that is updated dynamically as new recessions are detected or the economy returns to expansion---defined by the indicator value returning to 0.

Formally, a classifier $k$ is made up of an indicator $i(t) \geq 0$ and threshold $\z >0$. The classifier keeps track of the state of the economy: $r(k,t) = 0$ if the economy is in expansion at time $t$ and $r(k,t) = 1$ if the economy is in recession at time $t$.
Initially, the economy is not in recession ($r(k,0) = 0$). In period $t$, if the economy was previously in recession ($r(k,t-1) = 1$), it remains in recession if the indicator is positive ($r(k,t) = 1$ if $i(t)>0$), but it enters an expansion if the indicator falls to 0 ($r(k,t) = 0$ if $i(t)=0$). If the economy was previously in expansion ($r(k,t-1) = 0$), it remains in expansion if the indicator is below the threshold ($r(k,t) = 0$ if $i(t)<\z$), but it enters a recession if the indicator crosses the threshold ($r(k,t) = 1$ if $i(t)\geq \z$). If the classifier moves from expansion to recession in period $t$ for the $j$th time, the detection date of recession $j$ is set to $t$: $d(k,j) = t$.

A simpler way to proceed would have been to delineate recessions simply as periods when the indicator is above the threshold. This is for example the approach taken by \citet{S19} and \citet{MS25}. The simple approach is not entirely desirable because it creates noise when the economy is exiting recessions. As the recession ends, the unemployment rate increases less rapidly and the vacancy rate declines less rapidly, so indicators start falling toward the threshold. During their fall, indicators sometimes drop below the threshold, then temporarily climb above the threshold, before falling down below it again. (Such blips appear on the indicator in figure~\ref{f:errors} in the aftermath of the 1990 and 2001 recessions.) This secondary period above the threshold is not a new recession, however, and it should not be counted as such. 

To eliminate such false positives, created by changes in the pace of economic slowdown around the threshold, I require that the economy is in expansion before entering a new recession, and I require that the indicator falls to 0 for the economy to be in expansion. This classification procedure adds a bit of computational complexity, but it provides a more logical and more robust classification.

\begin{figure}[t]
\subcaptionbox{False negatives with high threshold \label{f:fn}}{\smallgraphic[4]{\pdf}}\hfill
\subcaptionbox{False positives with low threshold \label{f:fp}}{\smallgraphic[5]{\pdf}}\\
\subcaptionbox{Perfect probabilistic classifier \label{f:perfect}}{\smallgraphic[6]{\pdf}}\hfill
\subcaptionbox{Mistimed recession detection \label{f:mistimed}}{\smallgraphic[7]{\pdf}}
\caption{Possible classification errors}
\label{f:errors}\end{figure}

\subsection{Selecting perfect classifiers}

I select perfect classifiers in two stages. The first stage counts detection signals. For each recession indicator, I divide its amplitude over the training period (from 0 to its maximum) into 10,000 equal steps, and I consider every threshold on that grid. In total, I therefore evaluate 95,832 $\times$ 10,000 $=$ 958,320,000 single-threshold classifiers. For each of them, I compute the number of recessions detected between April 1929 and December 2025. The evaluation starts in April 1929 because this is when the data become available. It stops in December 2025 because it is too early to say if and when a recession occurred after that. Between April 1929 and December 2025, the NBER identifies $J = 15$ recessions. I therefore only keep the thresholds that produce exactly 15 detection signals during that time. When the threshold is too high, the classifier produces fewer signals, because it misses some recessions: false negatives (figure~\ref{f:fn}). When the threshold is too low, the classifier produces more signals, because it detects nonexistent recessions: false positives (figure~\ref{f:fp}).

For a given indicator, the thresholds that produce exactly 15 signals form an interval. The reason is that raising the threshold can only eliminate detection signals: since the classifier only returns to expansion when the indicator falls back to zero, any episode during which the indicator crosses a high threshold is also an episode during which it crosses a lower threshold. I denote by $\z_l$ the bottom of the interval and by $\z_h$ its top. Pairing the two thresholds yields a dual-threshold classifier $k$. The classifier issues an early warning for recession $j$ when the indicator crosses $\z_l$, at date $d_l(k,j)$; it issues a final warning for recession $j$ when the indicator crosses $\z_h$, at date $d_h(k,j)$. As shown in section~\ref{s:probability}, the early warning is when the recession probability becomes positive, and the final warning is when the recession probability reaches 1.

The second stage checks the timing of the warnings. Producing 15 signals is not sufficient to detect the 15 recessions: a classifier could miss one recession and pick up another event, or it could detect a recession long before it started. I therefore impose two timing conditions. Let $s(j)$ and $e(j)$ denote the start and end dates of recession $j$, as determined by the NBER. First, the early warning for each recession must occur between the end of the previous recession and the end of the current recession:
\begin{equation}
e(j-1) \leq d_l(k,j) \leq e(j),
\label{e:early}\end{equation}
where $e(0)$ is the start of the training period. The recession cannot be detected before the previous recession is over, and it must be detected before it is over itself. Second, the final warning for each recession must occur between the start of the current recession and the start of the next recession:
\begin{equation}
s(j) \leq d_h(k,j) \leq s(j+1),
\label{e:final}\end{equation}
where $s(J+1)$ is the end of the training period. A recession cannot be certain before it has started, and it must be certain before the next recession starts. In practice, I impose the two conditions month by month over the training period: at any date, the number of early warnings issued so far must be at least the number of recessions that have ended and at most one more; and the number of final warnings issued so far must be at most the number of recessions that have started and at least one fewer.

Figure~\ref{f:errors} illustrates how the procedure works. The classifier in figure~\ref{f:perfect} is perfect, as the early and final warnings are all well positioned, indicating actual recessions. The classifier in figure~\ref{f:mistimed} is not perfect, although it does detect the correct number of events. Indeed, the second final warning arrives before the second recession has even started, and the third early warning arrives before the second recession has even ended. This is the algorithmic way to see that the classifier detects an event that is not a recession and misses the final actual recession.

A dual-threshold classifier that satisfies \eqref{e:early} and \eqref{e:final} is perfect. It detects each of the 15 recessions, without false positives or negatives, and each of its warnings pertains to the recession that it is meant to detect. Overall, the procedure produces 1,028 perfect classifiers over April 1929--December 2025.

Although the indicators are constructed by filtering the data in many ways, perfection is difficult to achieve. Only about 1\% of the indicators yield a perfect classifier. And none of the 4,356 indicators based on unemployment alone---obtained by setting $\d = 1$ in \eqref{e:uv}---yields a perfect classifier. However the unemployment rate is smoothed, compared to its past values, and transformed, it cannot detect the 15 recessions of 1929--2025. Hence, no unemployment-only rule, such as the Sahm rule, can be perfect over the training period. The flexibility of the filtering is therefore not sufficient to produce perfect classifiers from any data source. 

Perfect classifiers only appear once unemployment is combined with vacancies. The reason is that the unemployment-vacancy combination depicts business cycles much better than unemployment alone---although of course not every combination works.

\subsection{Recession probability}\label{s:probability}

Each perfect classifier comes with two detection thresholds, $\z_l < \z_h$: a dual-threshold rule. The recession probability produced by classifier $k$ is built in two steps. The first step locates the indicator $i(t)$ in the $\z_l$--$\z_h$ band:
\begin{equation}
p(k,t) = \min{\max{\frac{i(t) -\z_l}{\z_h-\z_l},0},1}.
\label{e:p}\end{equation}
The location $p(k,t)$ is 0 when the indicator is below $\z_l$, 1 when the indicator is above $\z_h$, and it increases linearly with the indicator in between.

The second step recognizes that the probability that a recession has started cannot decline once the indicator has crossed $\z_l$, until the economy returns to an expansionary state. The probability therefore tracks the highest location reached by the indicator since it last was at zero:
\begin{equation}
P(k,t) = \begin{cases}
\max{P(k,t-1),\, p(k,t)} & \text{if } i(t) > 0\\
0 & \text{if } i(t) = 0.
\end{cases}
\label{e:pkt}\end{equation}
The probability is 0 as long as the indicator stays below the lower threshold. When the indicator first crosses the lower threshold, the classifier issues an early warning, and the probability becomes positive. The probability then rises whenever the indicator reaches a new high within the band, and it reaches 1 once the indicator crosses the upper threshold. If the indicator dips after having risen, the probability stays put. Finally, when the indicator returns to zero---which is when the recessions detected by the classifiers end---the economy is back in an expansionary state and the probability resets to 0.

The recession probability is a byproduct of our ignorance, itself caused by a dearth of macroeconomic data. The presumption is that there is a unique threshold separating recessions from nonrecessions. When the indicator crosses the true threshold from below, the economy experiences a recession. The challenge is that there is not enough data to identify this true threshold with exactitude. The true threshold is above $\z_l$ because the indicator has reached values up to $\z_l$ outside of recessions without triggering a recession. The true threshold is below $\z_h$ because any threshold above $\z_h$ would miss at least one recession. So the true threshold must be between $\z_l$ and $\z_h$. The range cannot be narrowed further without observing more recessions. Assuming that this unobservable threshold is uniformly distributed over $\z_l$--$\z_h$---a convenient and neutral assumption---the probability that a recession has started is the probability that the indicator has crossed the latent threshold at some point since the economy was last in an expansionary state. This is the probability that the latent threshold lies below the highest value reached by the indicator since it was last at zero, which is given by \eqref{e:pkt}. This is why the probability cannot decline during an episode: once the indicator has reached a certain level, the latent threshold has been crossed with the corresponding probability, whatever the indicator does next.

When several classifiers are used together, the recession probability is the average of the probabilities produced by the $K$ individual classifiers:
\begin{equation}
P(t) = \frac{1}{K} \cdot \sum_{k = 1}^{K} P(k,t).
\label{e:pt}\end{equation}
I aggregate the individual probabilities using equal weights for transparency.\footnote{In principle, the probabilities could be aggregated using generalized least-squares weights to correct for correlation across recession indicators. In practice, this approach is unreliable because the indicators are constructed from transformations of only two data series (the unemployment rate and the vacancy rate), so they are nearly collinear. Technically, the covariance matrix of the underlying indicators is nearly singular, so the generalized least-squares weights are numerically unstable.}

\subsection{Qualities of perfect classifiers}

Given that there are so many perfect classifiers, I need to describe them in some way to select the classifiers used for detection. Standard evaluation methods, like ROC curves, rank classifiers based on the number of false positives and negatives that they generate.\footnote{See for instance \citet[section~5.1.3]{M22a} and \citet[section~18.2.3]{P20}.} They are unsuitable here because the perfect classifiers generate no errors in sample. Instead, I evaluate each classifier $k$ on four qualities: earliness, finality, robustness, and accuracy. 

\paragraph{Detection errors} For each recession $j$, I compute the detection error of the early warning, which is the difference between the date when the recession officially started, $s(j)$, and the date when the classifier issued the early warning, $d_l(k,j)$:
\begin{equation}
\e(k,j) = s(j) - d_l(k,j).
\label{e:delay}\end{equation}
This is doable because each perfect classifier issues exactly one early warning per recession, so there are as many start dates as early warnings. If $\e(k,j)>0$, classifier $k$ detects recession $j$ with some anticipation. If instead $\e(k,j)<0$, classifier $k$ detects recession $j$ with some delay. The detection error of the final warning is computed in the same way, from $d_h(k,j)$ instead of $d_l(k,j)$:
\begin{equation}
\e_h(k,j) = s(j) - d_h(k,j).
\label{e:delayh}\end{equation}

\paragraph{Earliness} The earliness of classifier $k$ is measured by the mean detection error of its early warnings on the training sample:
\begin{equation}
\m(k) = \frac{1}{J} \cdot \sum_{j=1}^{J} \e(k,j),
\label{e:mu}\end{equation}
where $J$ is the number of recessions in the training sample. A classifier with higher earliness (higher $\m(k)$) gives a possible recession signal sooner, which allows policymakers, firms, workers, investors, and consumers to prepare for the recession.

\paragraph{Finality} The finality of classifier $k$ is measured by the mean detection error of its final warnings on the training sample:
\begin{equation}
\phi(k) = \frac{1}{J} \cdot \sum_{j=1}^{J} \e_h(k,j).
\label{e:phi}\end{equation}
A classifier with higher finality (higher $\phi(k)$) reaches certainty that a recession has started sooner. Because the final warning comes when the indicator crosses the upper threshold, finality depends on how low the upper threshold can be set while still producing a correct final warning in every recession.

\paragraph{Robustness}  The robustness of classifier $k$ is measured by the width of its threshold band:
\begin{equation}
R(k) = \frac{\z_h - \z_l}{\max_t i(t)},
\label{e:robustness}\end{equation}
where the maximum is taken over the training period. $R(k)$ is the share of the indicator's amplitude over the training period that is covered by the threshold band. Robustness measures how different the indicator is in recessionary and nonrecessionary states: the lower threshold is set by the highest values that the indicator reaches outside recessions, and the upper threshold by the lowest peak that the indicator reaches during recessions. Methods based on robustness are included because classifiers with wider margins between the two classes are expected to make fewer mistakes on new data \citep{STBWA98,B98}. Hence, even a classifier selected among many candidates can be expected to perform well out of sample if its margin is wide.\footnote{The idea of choosing the boundary with the widest margin goes back to \citet{VL63} and underlies support vector machines \citep{BGV92,CV95}. See \citet{BBL05} and \citet[chapter 12]{HTF09} for surveys.} 

\paragraph{Accuracy} Finally, the accuracy of classifier $k$ is measured by the standard deviation of the detection errors of its early warnings on the training sample:
\begin{equation}
\sigma(k) = \sqrt{\frac{1}{J} \cdot \sum_{j=1}^{J} \bs{\e(k,j) - \m(k)}^2}.
\label{e:sigma}\end{equation}
A classifier with higher accuracy (lower $\sigma(k)$) issues its early warnings at a more consistent distance from recession starts, so the timing of its early warnings is more predictable.

\section{Selecting classifiers from pseudo-out-of-sample tests}\label{s:learning}

Every perfect classifier fits the training data equally well, so the training data alone cannot determine which classifiers to use for detection. This section describes how pseudo-out-of-sample tests settle the question. I consider many ways of selecting perfect classifiers, based on the four qualities described in the previous section. None of the four qualities has an obvious claim to be the right one. Earliness and finality are the qualities that matter for a policymaker, but they are measured in sample. Robustness and accuracy measure how regularly the indicator behaves across recessions, which might matter out of sample. I apply each method to past episodes that were not in their training data, and keep the method that performs best on these pseudo-out-of-sample tests. That method is then applied to the full data available, and the classifier that it selects is used to detect recessions.

\subsection{Classifier ensembles}

When a method selects several classifiers, their recession probabilities are averaged according to \eqref{e:pt}, so the method produces an ensemble. Ensembles of several classifiers are included because ensembles are generally expected to perform better out of sample than single classifiers. Every perfect classifier fits the training data equally well, so which of them will be most accurate on new data is unknown in advance \citep{LW94,T03}; and when the classifiers disagree, the error of their average is smaller than the average of their errors \citep{KV95}. The same arguments underlie the early results on combining classifiers by voting, bagging, and boosting \citep{HS90,S90,B96a}.\footnote{\citet{D00}, \citet{Z12}, and \citet{SR18} survey this literature.} 

\subsection{Selection methods}

I consider 34 selection methods, built from the four qualities defined in section~\ref{s:ensemble}. The first 28 methods select the $K$ perfect classifiers with the highest earliness, finality, robustness, or accuracy, with $K = 1, 3, 5, 7, 10, 15, 20$. I label these methods by the initial of the quality and the number of classifiers: E1, E3, \ldots, E20 for earliness; F1, F3, \ldots, F20 for finality; R1, R3, \ldots, R20 for robustness; and A1, A3, \ldots, A20 for accuracy. The last 6 methods select the classifiers on the frontier of a pair of qualities: the perfect classifiers such that no other perfect classifier is at least as good on both qualities and strictly better on one. I label these methods by the initials of the two qualities: EF, ER, EA, FR, FA, and RA. For instance, the ER method selects the classifiers on the earliness-robustness frontier.

\subsection{Pseudo-out-of-sample tests}

A pseudo-out-of-sample test splits the data at a cutoff date. The test pretends that the data end in the December before the cutoff: the perfect classifiers are recomputed on that training period, and each of the 34 methods selects its classifiers among them. The selected classifiers are then scored on the next episode after the cutoff, which the training data do not contain.

I use 7 test windows: 1973--1977, 1978--1984, 1985--1994, 1995--2004, 2005--2014, 2015--2021, and 2022--2025. The 1973--1977 test is trained on data up to December 1972 and scored on the 1973--1975 recession. The next 5 tests are scored on the 6 recessions that the NBER Business Cycle Dating Committee has faced since its inception in 1978: the twin Volcker recessions for the 1978--1984 test, the Gulf War recession for the 1985--1994 test, the dot-com recession for the 1995--2004 test, the Great Recession for the 2005--2014 test, and the pandemic recession for the 2015--2021 test. The 2022--2025 test is scored on the 2023--2024 slowdown, which was not a recession but which several recession detection rules, including the Sahm and Michez rules, took for one \citep{DGN26}. Each window after 1977 is also an out-of-sample test in section~\ref{s:simulation}; the 1973--1977 window serves only as a pseudo-out-of-sample test.

The cutoffs fall roughly midway between recessions, so that no training period ends inside a recession. December 1972 falls between the 1969--1970 and 1973--1975 recessions; December 1977, between the 1973--1975 recession and the Volcker recessions; December 1984, between the Volcker recessions and the Gulf War recession; December 1994, between the Gulf War recession and the dot-com recession; December 2004, between the dot-com recession and the Great Recession; December 2014, between the Great Recession and the pandemic recession; and December 2021, between the pandemic recession and the 2023--2024 slowdown. 

Performance in each test is measured in probability-months, using the recession probability $P(t)$ given by the selected classifiers. The score only looks at recession months: it is the shortfall of the recession probability from 1 during the recessions covered by the test. Let $\Rc$ denote the set of months that belong to these recessions. The score is
\begin{equation}
S = \sum_{t \in \Rc} \bs{1 - P(t)}.
\label{e:score}\end{equation}
In the 1978--1984 test, $\Rc$ contains the months of both twin recessions, February--July 1980 and August 1981--November 1982; in the other recession tests, $\Rc$ contains the months of a single recession. The score does not look at the months outside recessions. In particular, it does not penalize classifiers whose recession probability remains positive after a recession, or between the twin recessions, as the timing of recession ends is not what the algorithm aims to detect. 

In the 2022--2025 test, there is no recession, so the score is the sum of the recession probability over the two years of the slowdown, January 2023--December 2024:
\begin{equation}
S = \sum_{t = 01/2023}^{12/2024} P(t).
\label{e:score2022}\end{equation}
In all tests, a lower score is better.

\subsection{Choosing the selection method}

The tests are used to choose the selection method. A test can only be used once its episode is over. So at any date, the selection methods are ranked by their average score over the tests completed by that date. The best method is the method with the lowest average score. The best method is then applied to the full data available at that date, which yields the classifiers used to detect recessions going forward.

Because a test enters the ranking only once its episode is over, the method chosen at any date uses only data available at that date. This is what makes the simulation of section~\ref{s:simulation} out of sample. That section simulates the algorithm at each of the episodes faced by the NBER Business Cycle Dating Committee since 1978. Section~\ref{s:current} describes the algorithm that results from all the tests available today.

\section{Simulating the algorithm over the lifetime of the NBER Business Cycle Dating Committee}\label{s:simulation}

This section puts the algorithm in the shoes of the NBER Business Cycle Dating Committee. The committee has faced 6 recessionary episodes since its creation in 1978: the twin Volcker recessions of 1980 and 1981--1982, the Gulf War recession of 1990--1991, the dot-com recession of 2001, the Great Recession of 2008--2009, the pandemic recession of 2020, and the 2023--2024 labor market slowdown. In each out-of-sample test, everything the algorithm uses (perfect classifiers, test scores, and the ranking of methods) comes from data available at the start of the test, so the episode that follows is out of sample. 

The simulation consists of 6 out-of-sample tests, one per episode, starting in January 1978, 1985, 1995, 2005, 2015, and 2022. At the start of each test, the algorithm ranks the 34 selection methods by their average score on the pseudo-out-of-sample tests completed so far, as described in section~\ref{s:learning}. The first out-of-sample test has a single completed pseudo-out-of-sample test, the 1973--1977 test; each subsequent out-of-sample test adds the window of the previous one. The algorithm then applies the best method to all the data available, from April 1929 to the previous December, which yields a classifier. Last, the algorithm uses this classifier to compute the recession probability month by month as new data come in.

\begin{sidewaystable}[p]
\caption{Decisions of the algorithm for out-of-sample tests}
\begin{tabular*}{\textwidth}{@{\extracolsep\fill}ccl*{9}{c}}
\toprule
& & & & & & \multicolumn{3}{c}{Selected method} & \multicolumn{2}{c}{Runner-up method} & \\
\cmidrule{6-8}\cmidrule{9-10}
\multicolumn{2}{c}{Test dates} & & Training & Training & Perfect & & Classifiers & Average & & Average & Score on \\
\cmidrule{1-2}
Start & End & Episode & period & recessions & classifiers & Method & selected & score & Method & score & next episode \\
\midrule
Jan 1978 & Dec 1984 & Twin Volcker recessions & 1929--1977 & 9 & 5,797 & R1 & 1 & 8.95 & ER & 9.08 & 7.94 \\
Jan 1985 & Dec 1994& Gulf War recession & 1929--1984 & 11 & 1,974 & F5 & 5 & 6.00 & F3 & 6.01 & 1.54 \\
Jan 1995& Dec 2004& Dot-com recession & 1929--1994 & 12 & 1,930 & F1 & 1 & 4.10 & F3 & 4.40 & 1.35 \\
Jan 2005& Dec 2014& Great Recession & 1929--2004 & 13 & 1,917 & F1 & 1 & 3.41 & F3 & 3.69 & 4.25 \\
Jan 2015& Dec 2021& Pandemic recession & 1929--2014 & 14 & 1,917 & F1 & 1 & 3.58 & F3 & 3.94 & 1.13 \\
Jan 2022& Dec 2025& 2023--2024 slowdown & 1929--2021 & 15 & 1,521 & F1 & 1 & 3.17 & F3 & 3.49 & 6.96 \\
\midrule
Jan 2026 & Aug 2026 & Current & 1929--2025 & 15 & 1,028 & F1 & 1 & 3.71 & F3 & 3.83 & -- \\
\bottomrule
\end{tabular*}
\note{In each out-of-sample test, the training period runs from April 1929 to the December before the test. The perfect classifiers are the dual-threshold classifiers that satisfy \eqref{e:early} and \eqref{e:final} over the training period. The average score of a selection method is its average score, in probability-months, on the pseudo-out-of-sample tests completed before the test: the 1973--1977 test for the 1978--1984 test; the 1973--1977 and 1978--1984 tests for the 1985--1994 test; and so on, up to all 7 tests for the current algorithm. The scores are given by \eqref{e:score} and \eqref{e:score2022}; a lower score is better. The score on the next episode is the score of the selected method in the pseudo-out-of-sample test on the same window. The selection methods are described in section~\ref{s:learning}.}
\label{t:committee}\end{sidewaystable}

\begin{sidewaystable}[p]
\caption{Classifiers selected by the algorithm for out-of-sample tests}
\begin{tabular*}{\textwidth}{@{\extracolsep\fill}ccl*{10}{c}}
\toprule
& & & & & & & & & \multicolumn{4}{c}{In-sample warning delay (months)}\\
\cmidrule{9-12}
& & & & & & & & & \multicolumn{2}{c}{Training period}& \multicolumn{2}{c}{Since 1978}\\
\cmidrule{9-10}\cmidrule{11-12}
\multicolumn{2}{c}{Test dates} & & Smoothing method & Turning & Box-Cox & Combination & Thresholds & Robustness & & & & \\
\cmidrule{1-2}
Start & End & Method & \& parameter $\a$ & horizon $\b$ & parameter $\g$ & \& weight $\d$ & ($\times 100$)& (\%) & Early & Final & Early & Final\\
\midrule
Jan 1978 & Dec 1984 & R1 & simple, 11 & 2m & 0.3 & min-max, 1 & 0.36--1.82 & 35.5 & 3.9 & 6.7 & -- & -- \\
\addlinespace
Jan 1985 & Dec 1994& F5 & exponential, 0.2 & 2m & 1 & min-max, 1 & 0.066--0.118 & 11.8 & 2.4 & 3.5 & 0.5 & 1.0 \\
& & exponential, 0.2 & 2m & 0.9 & min-max, 1 & 0.093--0.170 & 13.3 & 2.4 & 3.6 & 0.5 & 1.0 \\
& & exponential, 0.3 & 3m & 0.2 & u-v, 0 & 4.11--4.18 & 0.6 & 3.6 & 3.6 & 3.0 & 3.0 \\
& & exponential, 0.7 & 7m & 1 & min-max, 1 & 0.35--0.51 & 6.8 & 2.0 & 3.7 & 0.5 & 2.0 \\
& & exponential, 0.5 & 4m & 0.1 & u-v, 0.2 & 8.16--9.31 & 4.4 & 3.5 & 3.7 & 3.0 & 3.0 \\
\addlinespace
Jan 1995& Dec 2004& F1 & exponential, 0.2 & 2m & 1 & min-max, 1 & 0.066--0.118 & 11.8 & 2.2 & 3.3 & 0.3 & 1.0 \\
Jan 2005& Dec 2014& F1 & exponential, 0.2 & 2m & 1 & min-max, 1 & 0.066--0.118 & 11.8 & 2.0 & 3.3 & 0.3 & 1.5 \\
Jan 2015& Dec 2021& F1 & exponential, 0.2 & 2m & 1 & min-max, 1 & 0.066--0.118 & 11.8 & 2.1 & 3.6 & 0.8 & 2.6 \\
Jan 2022& Dec 2025& F1 & exponential, 0.2 & 2m & 1 & min-max, 1 & 0.066--0.118 & 11.8 & 2.0 & 3.5 & 0.8 & 2.5 \\
\midrule
Jan 2026 & Aug 2026 & F1 & exponential, 0.7 & 8m & 1 & min-max, 1 & 0.47--0.53 & 2.4 & 3.0 & 3.7 & 2.2 & 2.8 \\
\bottomrule
\end{tabular*}
\note{In the 1985--1994 test, the F5 method selects the 5 classifiers with highest finality, which are listed from highest to lowest finality; in the other tests, the selected method selects a single classifier. The simple smoothing method with parameter $\a$ is given by \eqref{e:usma} and \eqref{e:vsma}; the exponential smoothing method with parameter $\a$ is given by \eqref{e:uema} and \eqref{e:vema}. The turning horizon $\b$ enters the construction of the classifiers through \eqref{e:umin} and \eqref{e:vmax}. The Box-Cox parameter $\g$ enters the construction of the classifiers through \eqref{e:uboxcox} and \eqref{e:vboxcox}. The u-v combination method with weight $\d$ is given by \eqref{e:uv}; the min-max combination method with weight $\d$ is given by \eqref{e:minmax}. The thresholds are expressed in the indicator's original units and multiplied by 100 for clarity; they are in percentage points when $\g = 1$. Robustness is given by \eqref{e:robustness}. The in-sample warning delays are minus the means of the detection errors \eqref{e:delay} and \eqref{e:delayh}; a positive delay means that the warnings lag the recession onset on average.}
\label{t:classifiers}\end{sidewaystable}

\begin{sidewaystable}[p]
\caption{Performance of the algorithm in out-of-sample tests}
\begin{tabular*}{\textwidth}{@{\extracolsep\fill}ccl*{8}{c}}
\toprule
& & & & \multicolumn{2}{c}{Early warning} & \multicolumn{2}{c}{Final warning} & \multicolumn{3}{c}{NBER announcement} \\
\cmidrule{5-6}\cmidrule{7-8}\cmidrule{9-11}
\multicolumn{2}{c}{Test dates} & \multicolumn{2}{c}{Episode}  & & Months & & Months & & Months after & Months after \\
\cmidrule{1-2}\cmidrule{3-4}
Start & End & Name & Onset & Date & after onset & Date & after onset & Date & early warning & final warning \\
\midrule
Jan 1978 & Dec 1984 & First Volcker recession & Feb 1980 & Feb 1980 & 0 & Jun 1980 & 4 & Jun 1980 & 4 & 0 \\
Jan 1978 & Dec 1984 & Second Volcker recession & Aug 1981 & Nov 1981 & 3 & Jun 1982 & 10 & Jan 1982 & 2 & $-5$ \\
Jan 1985 & Dec 1994 & Gulf War recession & Aug 1990 & Aug 1990 & 0 & Nov 1990 & 3 & Apr 1991 & 8 & 5 \\
Jan 1995 & Dec 2004 & Dot-com recession & Apr 2001 & Apr 2001 & 0 & Jul 2001 & 3 & Nov 2001 & 7 & 4 \\
Jan 2005 &  Dec 2014 & Great Recession & Jan 2008 & Apr 2008 & 3 & Aug 2008 & 7 & Dec 2008 & 8 & 4 \\
Jan 2015 & Dec 2021 & Pandemic recession & Mar 2020 & Apr 2020 & 1 & May 2020 & 2 & Jun 2020 & 2 & 1 \\
Jan 2022 & Dec 2025 & 2023--2024 slowdown & -- & Oct 2023 & -- & Jul 2024 & -- & -- & -- & -- \\
\midrule
\multicolumn{4}{l}{Average delay over the 6 recessions} & & 1.2 & & 4.8 & & 5.2 & 1.5 \\
\bottomrule
\end{tabular*}
\note{In each out-of-sample test, the recession probability is given by \eqref{e:pkt} for the classifier selected at the start of the test, described in table~\ref{t:classifiers}; in the 1985--1994 test, it is the average \eqref{e:pt} of the probabilities given by the 5 selected classifiers. The early warning is the first month in which the recession probability is positive; the final warning is the first month in which it reaches 100\%. The NBER announcements are from table~\ref{t:nber}. The months after the early and final warnings measure how long the NBER announcement came after each warning; a negative value means that the announcement preceded the warning.}
\label{t:oos}\end{sidewaystable}

\subsection{Twin Volcker recessions}

\begin{figure}[t]
\subcaptionbox{Ranking on pseudo-out-of-sample test \label{f:rank1978}}{\smallgraphic[8]{\pdf}}\hfill
\subcaptionbox{Selected recession classifier \label{f:classifier1978}}{\smallgraphic[9]{\pdf}}\\
\subcaptionbox{In-sample recession probability \label{f:is1978}}{\smallgraphic[10]{\pdf}}\hfill
\subcaptionbox{Out-of-sample recession probability \label{f:oos1978}}{\smallgraphic[11]{\pdf}}
\caption{Algorithm in 1978--1984}
\note{Panel~A ranks the 34 selection methods by their score on the 1973--1977 test, in probability-months; a lower score is better. Panel~B displays the classifier selected by the best method, R1, on April 1929--December 1977 data: the thick, purple line is the recession indicator, and the thin, pink lines are the thresholds. Panel~C displays the recession probability given by the classifier over the training period, April 1929--December 1977. Panel~D displays the recession probability given by the classifier out of sample, January 1978--December 1984; the vertical, pink lines mark the NBER announcements of recession starts. The recession probability is given by \eqref{e:pkt}. Shaded areas indicate recessions dated by the \citet{NBER23}.}
\label{f:1978}\end{figure}

In January 1978, the algorithm has data from April 1929 to December 1977, which contain 9 recessions. With these data, it can complete a single pseudo-out-of-sample test: training on data up to December 1972 and scoring on the 1973--1975 recession. In this test, the best method is R1, which selects the single most robust classifier, with 8.95 probability-months (figure~\ref{f:rank1978}). It is followed by ER, with 9.08 probability-months, and FR, with 9.27. Methods based on accuracy do worst, with 12.7 to 13.0 probability-months. The spread across methods is narrow because every method detects the 16-month 1973--1975 recession late.

Applied to the 5,797 perfect classifiers of 1929--1977, the R1 method selects the classifier with the widest threshold band, which covers 35.5\% of the indicator's amplitude (figure~\ref{f:classifier1978} and table~\ref{t:classifiers}). The indicator is the minimum of unemployment and vacancy indicators, as in the Michez rule. But it smooths the data heavily, with a 12-month trailing average; it uses a 2-month turning-point horizon; and it is closer to percentage changes than to level changes ($\g = 0.3$). The price of the wide threshold band is a late final warning. In sample, the early warnings come on average 3.9 months after recession onset, but the final warnings only come after 6.7 months (figure~\ref{f:is1978}).

Out of sample, the classifier detects the first Volcker recession immediately (figure~\ref{f:oos1978}). The recession probability turns positive in February 1980, the month the recession started, at 7\%. It rises to 39\% in April and 95\% in May, and reaches 100\% in June 1980---the month in which the NBER announced the recession. The recession probability then remains at 100\% until May 1981 and returns to 0\% in June 1981, two months before the second recession started. For the second recession, which started in August 1981, the recession probability turns positive in November 1981, two months before the NBER announcement in January 1982, and passes 50\% in January 1982. It reaches 100\% only in June 1982, 5 months after the NBER announcement. 

The score of the classifier on the twin recessions is 7.9 probability-months. This is only the 26th score among the 34 methods in the 1978--1984 test, far behind the 2.0 probability-months of F1, F3, and F5 (table~\ref{t:scores}). The price of the wide threshold band appears again: the most robust classifier is slow to reach certainty that a recession has started.

The twin recessions are the most demanding test faced by perfect classifiers since 1929. Only 1,974 of the 5,797 classifiers that are perfect on data up to 1977---34\%---remain perfect once the twin recessions enter the training data. Most of the eliminated classifiers are based on indicators that do not return to 0 between the two recessions, so they cannot detect the second recession.\footnote{For this reason, the Michez indicator is not perfect over 1929--2025. The indicator does not return to 0 between the twin recessions, so the classifier does not return to expansion, and the 1981--1982 recession cannot be detected as a new recession. This was not an issue in \citet{MS25} because there a recession starts when the indicator crosses the threshold upward and ends when it crosses the threshold downward.}

\subsection{Gulf War recession}

\begin{figure}[t]
\subcaptionbox{Ranking on pseudo-out-of-sample tests \label{f:rank1985}}{\smallgraphic[12]{\pdf}}\hfill
\subcaptionbox{A classifier in the selected ensemble\label{f:classifier1985}}{\smallgraphic[13]{\pdf}}\\
\subcaptionbox{In-sample recession probability \label{f:is1985}}{\smallgraphic[14]{\pdf}}\hfill
\subcaptionbox{Out-of-sample recession probability \label{f:oos1985}}{\smallgraphic[15]{\pdf}}
\caption{Algorithm in 1985--1994}
\note{Panel~A ranks the 34 selection methods by their average score on the 1973--1977 and 1978--1984 tests, in probability-months; a lower score is better. Panel~B displays the first of the 5 classifiers selected by the best method, F5, on April 1929--December 1984 data (table~\ref{t:classifiers}): the thick, purple line is the recession indicator, and the thin, pink lines are the thresholds. Panel~C displays the recession probability given by the 5 classifiers over the training period, April 1929--December 1984. Panel~D displays the recession probability given by the 5 classifiers out of sample, January 1985--December 1994; the vertical, pink line marks the NBER announcement of the recession start. The thin, orange lines are the recession probabilities given by the individual classifiers, computed with \eqref{e:pkt}; the thick, purple line is their average, given by \eqref{e:pt}. Shaded areas indicate recessions dated by the \citet{NBER23}.}
\label{f:1985}\end{figure}

In January 1985, the algorithm has data up to December 1984, which contain 11 recessions, and two completed tests: the 1973--1977 test and the 1978--1984 test. Averaged over the two tests, the best method is F5, which selects the 5 classifiers with the highest finality, with 6.00 probability-months (figure~\ref{f:rank1985}). It is followed very closely by F3, with 6.01 probability-months, and F1, with 6.02, and then by F20, FR, F15, F7, EF, and F10: the 9 best methods all use finality. These methods take over because they reach certainty fast in both twin recessions: F1, F3, and F5 score 2.0 probability-months in the 1978--1984 test, the best score of all methods. R1, the winner in 1978, falls to 20th place; the methods based on robustness now occupy the last places.

Applied to the 1,974 perfect classifiers of 1929--1984, the F5 method selects 5 classifiers (figure~\ref{f:classifier1985} and table~\ref{t:classifiers}). All of them smooth the data with an exponentially weighted moving average and use a short turning-point horizon, of 2 to 7 months. Three of them are minimum indicators of level changes or near-level changes in unemployment and vacancy rates. The first of them, which has the highest finality, is the classifier that the algorithm selects from 1995 onward. The other two are linear combinations of unemployment and vacancy indicators that put most or all weight on vacancies and are close to percentage changes. In sample, the final warnings of the 5 classifiers come on average 3.5 to 3.7 months after recession onset (figure~\ref{f:is1985}).

Out of sample, the ensemble detects the Gulf War recession immediately (figure~\ref{f:oos1985}). The recession probability turns positive in August 1990, the month the recession started, at 29\%. It rises to 57\% in September and 60\% in October, and reaches 100\% in November 1990, 5 months before the NBER announcement in April 1991. The probability remains at 100\% until April 1992, then falls as some of the classifiers return to 0\%, and is back at 0\% in November 1992. The score of the ensemble on the Gulf War recession is 1.5 probability-months (table~\ref{t:scores}).

The Gulf War recession barely changes the set of perfect classifiers: 1,930 of the 1,974 perfect classifiers, or 98\%, remain perfect once it enters the training data.

\subsection{Dot-com recession}

\begin{figure}[t]
\subcaptionbox{Ranking on pseudo-out-of-sample tests \label{f:rank1995}}{\smallgraphic[16]{\pdf}}\hfill
\subcaptionbox{Selected recession classifier \label{f:classifier1995}}{\smallgraphic[17]{\pdf}}\\
\subcaptionbox{In-sample recession probability \label{f:is1995}}{\smallgraphic[18]{\pdf}}\hfill
\subcaptionbox{Out-of-sample recession probability \label{f:oos1995}}{\smallgraphic[19]{\pdf}}
\caption{Algorithm in 1995--2004}
\note{Panel~A ranks the 34 selection methods by their average score on the 1973--1977, 1978--1984, and 1985--1994 tests, in probability-months; a lower score is better. Panel~B displays the classifier selected by the best method, F1, on April 1929--December 1994 data: the thick, purple line is the recession indicator, and the thin, pink lines are the thresholds. Panel~C displays the recession probability given by the classifier over the training period, April 1929--December 1994. Panel~D displays the recession probability given by the classifier out of sample, January 1995--December 2004; the vertical, pink line marks the NBER announcement of the recession start. The recession probability is given by \eqref{e:pkt}. Shaded areas indicate recessions dated by the \citet{NBER23}.}
\label{f:1995}\end{figure}

In January 1995, the algorithm has data up to December 1994, which contain 12 recessions, and three completed tests. The best method is now F1, which selects the single classifier with the highest finality, with 4.10 probability-months (figure~\ref{f:rank1995}). It moves ahead of F3, with 4.40, and F5, with 4.52, thanks to its excellent performance on the Gulf War recession. The 10 best methods all use finality, while the methods based on robustness remain at the bottom of the ranking.

Applied to the 1,930 perfect classifiers of 1929--1994, the F1 method selects a minimum indicator of level changes in unemployment and vacancy rates, like the Michez indicator (figure~\ref{f:classifier1995}). It is the classifier with highest finality in the 1985 ensemble. It differs from the Michez indicator in its filtering: it smooths the data with an exponentially weighted moving average with $\a = 0.2$, and it uses a 2-month turning-point horizon. Its thresholds are 0.066pp and 0.118pp, and its threshold band covers 11.8\% of the indicator's amplitude. In sample, its early warnings come on average 2.2 months after recession onset, and its final warnings 3.3 months after onset (figure~\ref{f:is1995}).

Out of sample, the classifier detects the dot-com recession immediately (figure~\ref{f:oos1995}). The recession probability turns positive in April 2001, the month the recession started, at 46\%. It rises to 72\% in June 2001 and reaches 100\% in July 2001, four months before the NBER announcement in November 2001. The probability then remains at 100\% until June 2003, during the jobless recovery, and returns to 0\% in July 2003. The score of the classifier on the dot-com recession is 1.4 probability-months.

Again, the recession barely changes the set of perfect classifiers: 1,917 of the 1,930 perfect classifiers, or 99\%, remain perfect once the dot-com recession enters the training data.

\subsection{Great Recession}

\begin{figure}[t]
\subcaptionbox{Ranking on pseudo-out-of-sample tests \label{f:rank2005}}{\smallgraphic[20]{\pdf}}\hfill
\subcaptionbox{Selected recession classifier \label{f:classifier2005}}{\smallgraphic[21]{\pdf}}\\
\subcaptionbox{In-sample recession probability \label{f:is2005}}{\smallgraphic[22]{\pdf}}\hfill
\subcaptionbox{Out-of-sample recession probability \label{f:oos2005}}{\smallgraphic[23]{\pdf}}
\caption{Algorithm in 2005--2014}
\note{Panel~A ranks the 34 selection methods by their average score on the 1973--1977, 1978--1984, 1985--1994, and 1995--2004 tests, in probability-months; a lower score is better. Panel~B displays the classifier selected by the best method, F1, on April 1929--December 2004 data: the thick, purple line is the recession indicator, and the thin, pink lines are the thresholds. Panel~C displays the recession probability given by the classifier over the training period, April 1929--December 2004. Panel~D displays the recession probability given by the classifier out of sample, January 2005--December 2014; the vertical, pink line marks the NBER announcement of the recession start. The recession probability is given by \eqref{e:pkt}. Shaded areas indicate recessions dated by the \citet{NBER23}.}
\label{f:2005}\end{figure}

In January 2005, the algorithm has data up to December 2004, which contain 13 recessions, and four completed tests. F1 remains the best method, with 3.41 probability-months, ahead of F3, with 3.69, and F5, with 3.90 (figure~\ref{f:rank2005}).

The F1 method selects the same classifier as in 1995, with the same thresholds (figure~\ref{f:classifier2005}): the dot-com recession changes the set of perfect classifiers only marginally, and it does not change the classifier with highest finality. In sample, its early warnings come on average 2.0 months after recession onset, and its final warnings 3.3 months after onset (figure~\ref{f:is2005}).

Out of sample, the classifier detects the Great Recession, which started in January 2008, in April 2008, when the recession probability turns positive (figure~\ref{f:oos2005}). The probability reaches 100\% in August 2008, four months before the NBER announcement in December 2008. This is quite good because in the summer of 2008, it was not obvious that the recession had already started. In August 2008, \citet{L08} circulated a working paper through the NBER arguing that there was no chance that the economy was currently in recession---and mocking Warren Buffett and others who were arguing that the economy had indeed entered a recession. The probability returns to 0\% in February 2010. The score of the classifier on the Great Recession is 4.3 probability-months, the best of the 34 methods in the 2005--2014 test.

All 1,917 perfect classifiers remain perfect once the Great Recession enters the training data.

\subsection{Pandemic recession}

\begin{figure}[t]
\subcaptionbox{Ranking on pseudo-out-of-sample tests \label{f:rank2015}}{\smallgraphic[24]{\pdf}}\hfill
\subcaptionbox{Selected recession classifier \label{f:classifier2015}}{\smallgraphic[25]{\pdf}}\\
\subcaptionbox{In-sample recession probability \label{f:is2015}}{\smallgraphic[26]{\pdf}}\hfill
\subcaptionbox{Out-of-sample recession probability \label{f:oos2015}}{\smallgraphic[27]{\pdf}}
\caption{Algorithm in 2015--2021}
\note{Panel~A ranks the 34 selection methods by their average score on the 1973--1977, 1978--1984, 1985--1994, 1995--2004, and 2005--2014 tests, in probability-months; a lower score is better. Panel~B displays the classifier selected by the best method, F1, on April 1929--December 2014 data: the thick, purple line is the recession indicator, and the thin, pink lines are the thresholds. Panel~C displays the recession probability given by the classifier over the training period, April 1929--December 2014. Panel~D displays the recession probability given by the classifier out of sample, January 2015--December 2021; the vertical, pink line marks the NBER announcement of the recession start. The recession probability is given by \eqref{e:pkt}. Shaded areas indicate recessions dated by the \citet{NBER23}.}
\label{f:2015}\end{figure}

In January 2015, the algorithm has data up to December 2014, which contain 14 recessions, and 5 completed tests. F1 remains the best method, with 3.58 probability-months, ahead of F3, with 3.94, and F5, with 4.21 (figure~\ref{f:rank2015}).

The F1 method selects the same classifier as in 1995 and 2005, with the same thresholds (figure~\ref{f:classifier2015}): the Great Recession changes neither the set of perfect classifiers nor the classifier with highest finality. In sample, its early warnings come on average 2.1 months after recession onset, and its final warnings 3.6 months after onset (figure~\ref{f:is2015}).

Out of sample, the classifier detects the pandemic recession, which started in March 2020, in April 2020 (figure~\ref{f:oos2015}). The recession probability reaches 100\% in May 2020, one month before the NBER announcement in June 2020. It returns to 0\% in August 2020. The score of the classifier on the pandemic recession is 1.1 probability-months.

The pandemic recession is a sharper test than the three previous recessions: 1,521 of the 1,917 perfect classifiers, or 79\%, remain perfect once it enters the training data. The classifier selected since 1995 is among them.

\subsection{2023--2024 labor market slowdown}

\begin{figure}[t]
\subcaptionbox{Ranking on pseudo-out-of-sample tests \label{f:rank2022}}{\smallgraphic[28]{\pdf}}\hfill
\subcaptionbox{Selected recession classifier \label{f:classifier2022}}{\smallgraphic[29]{\pdf}}\\
\subcaptionbox{In-sample recession probability \label{f:is2022}}{\smallgraphic[30]{\pdf}}\hfill
\subcaptionbox{Out-of-sample recession probability \label{f:oos2022}}{\smallgraphic[31]{\pdf}}
\caption{Algorithm in 2022--2025}
\note{Panel~A ranks the 34 selection methods by their average score on the 1973--1977, 1978--1984, 1985--1994, 1995--2004, 2005--2014, and 2015--2021 tests, in probability-months; a lower score is better. Panel~B displays the classifier selected by the best method, F1, on April 1929--December 2021 data: the thick, purple line is the recession indicator, and the thin, pink lines are the thresholds. Panel~C displays the recession probability given by the classifier over the training period, April 1929--December 2021. Panel~D displays the recession probability given by the classifier out of sample, January 2022--December 2025. The recession probability is given by \eqref{e:pkt}. Shaded areas indicate recessions dated by the \citet{NBER23}.}
\label{f:2022}\end{figure}

In January 2022, the algorithm has data up to December 2021, which contain all 15 recessions since 1929, and 6 completed tests, which cover the seven recessions since 1973. F1 remains the best method, with 3.17 probability-months, ahead of F3, with 3.49, F5, F7, and EF (figure~\ref{f:rank2022}). This ranking is the ranking obtained from the recessions alone, since the slowdown has not yet happened. In that ranking, the 10 best methods all use finality, methods based on earliness are in the middle, and methods based on accuracy and robustness are at the bottom.

The F1 method selects again the same classifier, with the same thresholds (figure~\ref{f:classifier2022}). In sample, its early warnings come on average 2.0 months after recession onset, and its final warnings 3.5 months after onset (figure~\ref{f:is2022}).

Out of sample, the classifier takes the slowdown for a recession (figure~\ref{f:oos2022}). The recession probability turns positive in October 2023 and reaches 100\% in July 2024. It returns to 0\% in January 2025, and it turns positive again in September 2025, at 12\%, where it remains through December 2025. The score of the classifier on the slowdown is 7.0 probability-months. So the algorithm would have declared a recession in July 2024, which the NBER never did. 

The algorithm made the same mistake as the Sahm rule, Michez rule, and yield-curve rules, which all signaled a recession during the slowdown \citep{DGN26}. The mistake is understandable, as even product market data signaled a slowdown at the time: industrial production contracted between September 2023 and January 2024 and again between June 2024 and November 2024 \citep{INDPRO}. And the mistake was not specific to the classifier selected in 2022, which is also the classifier selected in 1995, 2005, and 2015. Had they been kept until then, the classifiers selected in 1985 would have given a recession probability of 75\% during the slowdown (figure~\ref{f:oos1985full}), and the classifier selected in 1978 a probability of 42\% (figure~\ref{f:oos1978full}).

The slowdown is the most informative episode in the labor market since the early 1980s. Only 1,028 of the 1,521 perfect classifiers, or 68\%, remain perfect once the slowdown enters the training data. The slowdown eliminates more classifiers than any of the four recessions that followed the twin recessions, although it was not a recession. Among the eliminated classifiers is the classifier selected in every out-of-sample test since 1995.

\section{Current state of the algorithm}\label{s:current}

\begin{figure}[t]
\subcaptionbox{Ranking on pseudo-out-of-sample tests \label{f:rankcurrent}}{\smallgraphic[32]{\pdf}}\hfill
\subcaptionbox{Selected recession classifier \label{f:classifiercurrent}}{\smallgraphic[33]{\pdf}}\\
\subcaptionbox{In-sample recession probability \label{f:iscurrent}}{\smallgraphic[34]{\pdf}}\hfill
\subcaptionbox{Out-of-sample recession probability \label{f:ooscurrent}}{\smallgraphic[35]{\pdf}}
\caption{Current state of the algorithm}
\note{Panel~A ranks the 34 selection methods by their average score on the 7 pseudo-out-of-sample tests, in probability-months; a lower score is better. Panel~B displays the classifier selected by the best method, F1, on April 1929--December 2025 data: the thick, purple line is the recession indicator, and the thin, pink lines are the thresholds. Panel~C displays the recession probability given by the classifier over the training period, April 1929--December 2025. Panel~D displays the recession probability given by the classifier out of sample, January--August 2026. The recession probability is given by \eqref{e:pkt}. Shaded areas indicate recessions dated by the \citet{NBER23}.}
\label{f:current}\end{figure}

The current algorithm has data up to December 2025 and all 7 pseudo-out-of-sample tests, including the 2022--2025 test scored on the slowdown. It selects its method based on all 7 tests and applies it to the 1929--2025 data.

\subsection{Ranking of selection methods}

\begin{table}[p]
\caption{Scores of the selection methods in the pseudo-out-of-sample tests}
\begin{tabular*}{\textwidth}{@{\extracolsep{\fill}}l*{9}{c}}
\toprule
Start:&Jan 1973&Jan 1978 & Jan 1985 & Jan 1995 & Jan 2005 & Jan 2015 & Jan 2022&& \\
End:& Dec 1977 & Dec 1984 & Dec 1994 & Dec 2004 & Dec 2014 & Dec 2021 & Dec 2025 & Average & Rank \\
\midrule
E1 & 9.89 & 6.70 & 3.88 & 3.14 & 7.01 & 2.00 & 22.98 & 7.94 & 34 \\
E3 & 9.73 & 5.14 & 3.11 & 3.19 & 7.03 & 2.00 & 22.01 & 7.46 & 33 \\
E5 & 10.03 & 7.17 & 2.55 & 3.07 & 7.02 & 2.00 & 15.06 & 6.70 & 31 \\
E7 & 10.13 & 7.98 & 2.30 & 3.03 & 6.82 & 1.71 & 16.55 & 6.93 & 32 \\
E10 & 9.92 & 6.11 & 2.09 & 3.01 & 6.79 & 1.70 & 16.43 & 6.58 & 30 \\
E15 & 9.71 & 4.68 & 2.06 & 3.10 & 6.82 & 1.67 & 16.37 & 6.35 & 29 \\
E20 & 9.67 & 4.40 & 2.00 & 3.16 & 6.79 & 1.70 & 15.43 & 6.16 & 28 \\
\midrule
F1 & 10.04 & 2.00 & 0.25 & 1.35 & 4.25 & 1.13 & 6.96 & 3.71 & 1 \\
F3 & 10.01 & 2.00 & 1.19 & 1.55 & 4.97 & 1.21 & 5.86 & 3.83 & 2 \\
F5 & 10.01 & 2.00 & 1.54 & 2.07 & 5.44 & 1.24 & 5.30 & 3.94 & 3 \\
F7 & 10.01 & 2.61 & 1.96 & 2.46 & 5.63 & 1.24 & 4.89 & 4.11 & 4 \\
F10 & 10.04 & 2.73 & 1.85 & 3.18 & 5.74 & 1.32 & 4.63 & 4.21 & 8 \\
F15 & 10.16 & 2.40 & 2.01 & 2.85 & 5.64 & 1.38 & 4.68 & 4.16 & 6 \\
F20 & 10.09 & 2.30 & 1.72 & 3.16 & 5.83 & 1.41 & 4.33 & 4.12 & 5 \\
\midrule
R1 & 8.95 & 7.94 & 4.45 & 4.22 & 6.36 & 1.85 & 3.43 & 5.31 & 19 \\
R3 & 9.70 & 10.11 & 4.93 & 4.25 & 6.51 & 1.85 & 3.54 & 5.84 & 22 \\
R5 & 9.68 & 9.99 & 5.28 & 4.23 & 6.59 & 1.85 & 3.48 & 5.87 & 24 \\
R7 & 9.82 & 10.23 & 5.32 & 4.35 & 6.76 & 1.85 & 3.48 & 5.97 & 27 \\
R10 & 9.65 & 9.81 & 5.19 & 4.38 & 6.78 & 1.85 & 3.42 & 5.87 & 23 \\
R15 & 9.75 & 10.09 & 5.19 & 4.42 & 6.81 & 1.83 & 3.37 & 5.92 & 25 \\
R20 & 9.79 & 10.14 & 5.17 & 4.46 & 6.89 & 1.83 & 3.40 & 5.95 & 26 \\
\midrule
A1 & 12.71 & 6.57 & 5.60 & 4.92 & 6.67 & 0.88 & 1.71 & 5.58 & 20 \\
A3 & 12.87 & 4.11 & 5.15 & 5.13 & 6.67 & 0.96 & 1.53 & 5.20 & 18 \\
A5 & 12.94 & 4.09 & 3.57 & 4.26 & 7.08 & 0.94 & 1.49 & 4.91 & 12 \\
A7 & 12.96 & 4.96 & 3.63 & 4.60 & 6.87 & 0.95 & 1.28 & 5.04 & 16 \\
A10 & 12.99 & 4.59 & 3.69 & 4.80 & 6.92 & 1.00 & 1.10 & 5.01 & 15 \\
A15 & 12.96 & 3.81 & 3.93 & 4.89 & 7.12 & 0.99 & 1.06 & 4.96 & 14 \\
A20 & 12.75 & 3.64 & 4.01 & 4.95 & 6.92 & 1.14 & 1.26 & 4.95 & 13 \\
\midrule
EF & 9.71 & 2.94 & 1.78 & 2.54 & 5.78 & 1.46 & 8.23 & 4.63 & 10 \\
ER & 9.08 & 5.18 & 3.42 & 3.45 & 6.36 & 1.66 & 7.05 & 5.17 & 17 \\
EA & 10.10 & 5.05 & 2.14 & 2.81 & 6.10 & 1.41 & 6.60 & 4.89 & 11 \\
FR & 9.27 & 3.20 & 2.33 & 2.97 & 5.88 & 1.51 & 6.83 & 4.57 & 9 \\
FA & 10.88 & 2.83 & 2.01 & 2.60 & 5.82 & 1.10 & 4.02 & 4.18 & 7 \\
RA & 10.55 & 8.23 & 4.80 & 4.33 & 7.07 & 1.70 & 2.67 & 5.62 & 21 \\
\bottomrule
\end{tabular*}
\note{The scores are in probability-months; a lower score is better. The score is given by \eqref{e:score} in the 1973--1977, 1978--1984, 1985--1994, 1995--2004, 2005--2014, and 2015--2021 tests, and by \eqref{e:score2022} in the 2022--2025 test. The average is the equal-weighted average across the 7 tests, and the rank is based on the average. Methods E1--E20, F1--F20, R1--R20, and A1--A20 select the 1--20 perfect classifiers with highest earliness, finality, robustness, and accuracy. Methods EF, ER, EA, FR, FA, and RA select the perfect classifiers on the corresponding quality frontiers.}
\label{t:scores}\end{table}

Averaged over the 7 tests, the best method is F1, which selects the single classifier with highest finality, with 3.71 probability-months (figure~\ref{f:rankcurrent} and table~\ref{t:scores}). It is followed by F3, with 3.83 probability-months, F5, with 3.94, F7, and F20. The 7 methods based on finality alone all rank in the top 8, together with the finality-accuracy frontier, FA. At the other end, the 7 methods based on earliness alone are the 7 worst methods, and E1, which selects the single classifier with highest earliness, is the worst of all. The slowdown is responsible for this collapse: E1 accumulates 23.0 probability-months of recession probability during the slowdown, more than any other method. Selecting classifiers for their in-sample earliness---which is how recession detection rules are usually designed---is therefore the worst way to perform in the pseudo-out-of-sample tests. The methods based on robustness alone rank from 19th to 27th, and the methods based on accuracy from 12th to 20th; accuracy-based methods move up the ranking because they handle the slowdown well. Averaging several classifiers with high finality does not improve on the single best one: F1 beats F3, F5, F7, F10, F15, and F20. No method performs well on all tests, however: F1 is behind R1 on the 1973--1977 test, and far behind the accuracy-based methods on the slowdown.

F1 is also the method that the algorithm uses from 1995 onward, and F5, which the algorithm uses in 1985, ranks third. So the choices made by the algorithm in real time are close to the choice made with all the tests available today. With hindsight, F1 would also have been a good choice in 1985: on the Gulf War recession, it scores 0.25 probability-months, against 1.5 for F5. But F1 was 0.02 probability-months behind F5 at the time.

\subsection{Selected classifier}

Applied to the 1,028 perfect classifiers of 1929--2025, the F1 method selects a minimum indicator of level changes in unemployment and vacancy rates (figure~\ref{f:classifiercurrent} and table~\ref{t:classifiers}). The selected classifier is a close relative of the Michez rule. Like the Michez rule, it takes the minimum of unemployment and vacancy indicators constructed from level changes ($\g = 1$). It differs from the Michez rule in its smoothing and turning-point horizon. The Michez rule uses a 3-month simple moving average ($\a = 2$) and a 12-month turning-point horizon ($\b = 12$). The selected classifier smooths the data more lightly, with an exponentially weighted moving average with $\a = 0.7$, and uses an 8-month turning-point horizon. With a shorter turning-point horizon, the indicator returns to zero sooner after a recession.

The key insight from the analysis by \citet{MS25} therefore remains valid: taking the minimum of unemployment and vacancy indicators provides better recession signals than relying on unemployment or vacancy indicators alone. The current classifier is a minimum indicator, and so are the classifiers selected in 1978 and from 1995 to 2022, as well as three of the 5 classifiers selected in 1985 (table~\ref{t:classifiers}). \citet{MS25} made this discovery by filtering labor market data in a specific way; but the result remains valid across a wide range of filtering.

The selection of this classifier shows how the algorithm corrects its own out-of-sample errors. The classifier selected from 1995 to 2022 is no longer perfect once the slowdown enters the training data: its indicator crossed its upper threshold during the slowdown, so no pair of thresholds can separate the slowdown from the 15 recessions. The current classifier was already perfect over 1929--2021, with the same upper threshold, 0.53pp, but with a lower threshold of 0.35pp; it ranked third by finality. Its indicator peaked within its threshold band during the slowdown, at 0.47pp. Once the slowdown enters the training data, the lower threshold rises to 0.47pp, so that the slowdown no longer triggers an early warning, and the robustness of the classifier falls from 7.4\% to 2.4\%. The algorithm thus learns from the slowdown, at the cost of a less robust classifier.

\subsection{Recession probability in sample}

Figure~\ref{f:iscurrent} shows the recession probability given by the selected classifier on the training sample, 1929--2025. In each of the 15 in-sample recessions, the recession probability rises shortly after the onset of the recession and reaches 1 before the recession ends---except in the pandemic recession, when it reaches 1 in May 2020, one month after the recession ended. Because the threshold band is narrow, the probability goes from 0 to 1 within at most two months; in 6 of the 15 recessions, it jumps from 0 to 1 in a single month. The probability returns to 0 around the end of the recession or some months later.\footnote{This algorithm does not aim to determine the ends of recessions. But it would be possible to build another algorithm on the same principles to determine recession ends as early as possible.}

On average the classifier detects recessions after their official start dates because the official start dates are backdated \citep{NBER21}. The NBER identifies recessions with hindsight, not in real time, so it is able to place the start dates slightly earlier than the classifier's detection dates. Over 1929--2025, the early warnings of the classifier occur on average 3.0 months after the recession onset, and its final warnings 3.7 months after the onset.

However, the classifier detects recessions significantly earlier than the NBER. On average, since its inception, the NBER Business Cycle Dating Committee has taken 6.3 months to announce recession starts (table~\ref{t:nber}). By comparison, over the same period, the early warnings of the classifier come on average 2.2 months after the recession onset, and the final warnings 2.8 months after the recession onset. So the early warnings precede the announcements of the NBER by 4.1 months, and even the final warnings precede them by 3.5 months. Out of sample, the final warnings of the simulated algorithm come later, 4.8 months after the recession onset on average (table~\ref{t:oos}), as should be expected: in sample, the classifier is selected knowing all the recessions.

\subsection{Recession probability out of sample}

Finally, I apply the classifier selected by the current algorithm to the period after its training period: January--August 2026. The aim is to assess whether a recession started after the end of the training period (figure~\ref{f:ooscurrent}). The classifier detects no recession start in 2026: it issues no early warning, so the recession probability remains at 0\% throughout the period. In fact, the recession probability has been at 0\% since December 2020, so the classifier gives no weight to the 2023--2024 slowdown either---which is by construction, since the classifier has been trained over a training period that includes the slowdown.

\section{Conclusion}

This section summarizes the paper's findings and draws a few lessons for recession detection and statistical classification more generally. 

\subsection{Summary}

The paper develops an algorithm for detecting recessions in real time that selects recession classifiers based on their pseudo-out-of-sample performance. To evaluate the algorithm, the paper simulates how the algorithm would have fared out of sample since 1978, when the NBER Business Cycle Dating Committee was created. Out of sample, the algorithm generates no false negatives: it detects the 6 recessions of the period. On average, the algorithm issues an early warning 5.2 months before the NBER announcement, and in 5 of the recessions, it issues a final warning before or in the month of the NBER announcement. But the algorithm generates a false positive: it takes the 2023--2024 labor market slowdown for a recession. 

Trained on all the data between April 1929 and December 2025, the algorithm uses the single recession classifier with the earliest final warnings. This classifier gives a 0\% recession probability between January and August 2026.

\subsection{Lessons for recession detection}

\paragraph{Unemployment and vacancies together describe the US business cycle far better than unemployment alone} With unemployment alone, it is impossible to produce a classifier that detects all 15 recessions since 1929 with correctly timed warnings. Most classifiers selected by the algorithm, including the current one, are minimum indicators of unemployment and vacancy, like the Michez rule developed by \citet{MS25}.

\paragraph{In-sample earliness is a poor guide to out-of-sample performance} The algorithm never selects classifiers for their earliness, and across all tests the 7 methods based on earliness are the 7 worst. What works best is to select the classifier that reaches certainty the fastest. As soon as two tests are available, the algorithm selects classifiers for their finality: F5 in 1985, and F1 from 1995 onward (table~\ref{t:classifiers}).

\paragraph{Recession classification has been remarkably stable during the lifetime of the NBER Business Cycle Dating Committee} Once the twin recessions of the early 1980s are in the training data, nearly all perfect classifiers remain perfect through the next three recessions: 1,974 classifiers are perfect on data up to 1984, and 1,917 of them are still perfect on data up to 2014. The choices of the algorithm are stable too: from 1995 to 2022, it keeps the same classifier, which detects the dot-com recession, the Great Recession, and the pandemic recession before the NBER announcements. 

\paragraph{Because recession classification is stable, retraining between recessions improves performance only subtly} A consequence of this stability is that the algorithm performs well even without retraining (appendix~\ref{a:retraining}). Whether it stops retraining in 1978, 1985, 1995, 2005, or 2015, it detects every subsequent recession, and its early warnings come before the NBER announcements by more than 5 months on average. Retraining mostly affects the speed at which the algorithm becomes certain that a recession has started. The classifiers selected with less data have wider threshold bands, so they reach certainty more slowly, but for the same reason they are less sensitive to the 2023--2024 slowdown.

\paragraph{There might be value in incorporating the unemployment and vacancy rates into the NBER Business Cycle Dating Committee's considerations} The committee does not rely on unemployment or vacancy data to date recessions. Instead, the committee relies mostly on ``real personal income less transfers, nonfarm payroll employment, employment as measured by the household survey, real personal consumption expenditures, wholesale-retail sales adjusted for price changes, and industrial production,'' and in recent decades it has put the most weight on the first two measures \citep{NBER25}. The \citet{NBER24} further argues that because the unemployment rate is trendless while the variables considered are growing, the unemployment rate cannot be used reliably to date business cycles. In fact, when the NBER announced the start of the dot-com recession, it explained that the unemployment rate was a lagging and noisy variable, so it was not appropriate to detect recessions \citep{NBER01}. When it announced the starts of the Great Recession and pandemic recession, it did not mention unemployment at all \citep{NBER08,NBER20}. Yet, the unemployment rate, when combined with the vacancy rate, produces early warnings well before the NBER announcements, and final warnings somewhat before (table~\ref{t:oos}).

\subsection{Lessons for statistical learning}

The simulation also challenges a few general principles of statistical learning: it seems that the classification of recessions does not obey common classification rules.

\paragraph{Recession classifiers with wide margins do not perform well in pseudo-out-of-sample tests} A first principle of classification is that classifiers with wider margins between the two classes generalize better. Here, the margin of a perfect classifier is its robustness, the width of the band between its two thresholds. Surprisingly, selecting classifiers for their robustness does not pay off: the most robust classifier, selected in 1978, is slow to reach certainty in the twin recessions, and across all tests the methods based on robustness alone rank from 19th to 27th out of 34. And the wide margin of the classifier selected from 1995 to 2022, which covers 11.8\% of its indicator's amplitude, did not protect it from the 2023--2024 slowdown, which it took for a recession with certainty.

\paragraph{Ensembles of recession classifiers do not perform well in pseudo-out-of-sample tests} A second principle of classification is that averaging several classifiers performs better out of sample than relying on a single one. Here, however, the algorithm chooses an ensemble only once, in 1985, when F5 beats F1 by 0.02 probability-months. The algorithm selects a single classifier in every other out-of-sample test (table~\ref{t:classifiers}). And on the Gulf War recession, the single classifier with highest finality would have done better than the ensemble of 5. One interpretation is that the gains from averaging require classifiers that make different errors, whereas perfect classifiers built from transformations of the same two series tend to make the same errors, so averaging mostly dilutes the best classifier with worse ones.

\subsection{Was there a recession in 2023--2024?}

The 2023--2024 slowdown was the most recession-like nonrecession experienced by the US labor market in the past half-century. The algorithm took it for a recession, as did most detection rules. The slowdown eliminated a third of the perfect classifiers, more than any recession since the early 1980s, and the current algorithm has learned from it. 

The 2023--2024 slowdown is an interesting case study: the economy cooled significantly without ever operating below full employment. The unemployment rate was below the full-employment rate of unemployment between 2022 and 2024, so the economy was operating above full employment \citep{ABCD}. At the same time, the economy cooled a lot: labor market tightness fell from 2 (very hot) to 1 (full employment) between 2022 and 2024 \citep{ABCD}. 

It is not implausible that the NBER committee opted to remain silent for this reason. Indeed, David Romer and Christina Romer, who are both members of the \citet{NBER26} Business Cycle Dating Committee, advocated in 2020 that the committee change its approach along those lines. In \citet{RR20}, they propose ``that the NBER modify its definition of a recession to emphasize rapid increases in the shortfall from normal rather than declines in economic activity.'' By this definition, the 2023--2024 slowdown would not be a recession since there was no shortfall from normal.\footnote{Nothing in the algorithm depends on the specific NBER chronology: any recession chronology could be used for training, including the new dates proposed by \citet{RR20}. A different chronology would change the training set, so different classifiers might be selected, but the procedure would work the same way.}

\bibliography{\bib}
\newpage
\appendix

\section{Performance of the algorithm without retraining}\label{a:retraining}

In the simulation of section~\ref{s:simulation}, the algorithm retrains at the start of every out-of-sample test: it re-ranks the selection methods, applies the best method to the data available, and uses the resulting classifier until the next test. This appendix asks how the algorithm would have performed had it stopped retraining and kept the classifiers selected at the start of one test until today, August 2026. Table~\ref{t:retraining} summarizes the results.

\begin{sidewaystable}[p]
\caption{Warnings issued by the algorithm without retraining}
\begin{tabular*}{\textwidth}{@{\extracolsep\fill}l*{8}{c}}
\toprule
& & & \multicolumn{2}{c}{Retrained algorithm} & \multicolumn{2}{c}{No retraining after 1978} & \multicolumn{2}{c}{No retraining after 1985} \\
\cmidrule{4-5}\cmidrule{6-7}\cmidrule{8-9}
& & NBER & Early & Final & Early & Final & Early & Final \\
Episode & Onset & announcement & warning & warning & warning & warning & warning & warning \\
\midrule
1980 recession & Feb 1980 & Jun 1980 & Feb 1980 & Jun 1980 & Feb 1980 & Jun 1980 & -- & -- \\
1981--1982 recession & Aug 1981 & Jan 1982 & Nov 1981 & Jun 1982 & Nov 1981 & Jun 1982 & -- & -- \\
Gulf War recession & Aug 1990 & Apr 1991 & Aug 1990 & Nov 1990 & Aug 1990 & Jun 1991 & Aug 1990 & Nov 1990 \\
Dot-com recession & Apr 2001 & Nov 2001 & Apr 2001 & Jul 2001 & Apr 2001 & Feb 2002 & Apr 2001 & Nov 2001 \\
Great Recession & Jan 2008 & Dec 2008 & Apr 2008 & Aug 2008 & Mar 2008 & Nov 2008 & Apr 2008 & Oct 2008 \\
Pandemic recession & Mar 2020 & Jun 2020 & Apr 2020 & May 2020 & Apr 2020 & Jun 2020 & Apr 2020 & May 2020 \\
\midrule
\multicolumn{3}{l}{Average delay after onset, 6 recessions (months)} & 1.2 & 4.8 & 1.0 & 7.8 & -- & -- \\
\multicolumn{3}{l}{Average delay after onset, 4 recessions since 1990 (months)} & 1.0 & 3.8 & 0.8 & 8.3 & 1.0 & 5.3 \\
\midrule
\multicolumn{3}{l}{Maximum recession probability in 2023--2025 (\%)} & \multicolumn{2}{c}{100} & \multicolumn{2}{c}{42} & \multicolumn{2}{c}{75} \\
\bottomrule
\end{tabular*}
\note{The retrained algorithm is the algorithm simulated in section~\ref{s:simulation}, whose warnings are reported in table~\ref{t:oos}; its recession probability is also the probability given without retraining after 1995, because the algorithm selects the same classifier from 1995 to 2022. Without retraining after 1978, the recession probability is given by \eqref{e:pkt} for the R1 classifier selected in 1978. Without retraining after 1985, the recession probability is the average \eqref{e:pt} of the probabilities given by the 5 classifiers selected by F5 in 1985. The early warning is the first month in which the recession probability is positive; the final warning is the first month in which it reaches 100\%. The delays are measured in months from the recession onset. The NBER announcements are from table~\ref{t:nber}.}
\label{t:retraining}\end{sidewaystable}

\subsection{No retraining after 1978} 

Kept unchanged, the R1 classifier selected in 1978 detects all 6 recessions of 1980--2020: it generates no false negatives (figure~\ref{f:oos1978full}). Its early warnings remain fast: they come 1.0 month after the recession onset on average, and 5.3 months before the NBER announcement, against 1.2 and 5.2 months for the retrained algorithm. For the Great Recession, its early warning even comes one month earlier than that of the retrained algorithm, in March 2008. But its final warnings are slow, because its threshold band is wide: they come 7.8 months after the recession onset on average, against 4.8 months for the retrained algorithm. In the Gulf War, dot-com, and 1981--1982 recessions, the probability reaches 100\% only after the NBER announcement---in June 1991, February 2002, and June 1982. 

During the 2023--2024 slowdown, the recession probability produced by the R1 classifier turns positive in October 2023 and rises to 42\% in July 2024. Apart from the 2023--2024 slowdown, there are no other false positives: the recession probability given by the 1978 classifier only turns positive during NBER recessions.

\subsection{No retraining after 1985} 

Kept unchanged, the ensemble of 5 classifiers selected by F5 in 1985 detects all four recessions of 1990--2020: the ensemble produces no false negatives (figure~\ref{f:oos1985full}). Its early warnings come 1.0 month after the recession onset on average, and 6.3 months before the NBER announcement. Its final warnings come 5.3 months after the recession onset on average, against 3.8 months for the retrained algorithm over the same four recessions. But they all come before the NBER announcement or in the same month: 5 months before in the Gulf War recession, in the same month in the dot-com recession, 2 months before in the Great Recession, and 1 month before in the pandemic recession. 

During the 2023--2024 slowdown, the recession probability produced by the ensemble turns positive in August 2023 and peaks at 75\% in August 2024. Apart from the slowdown, however, the ensemble generates no false positives: the recession probability given by the ensemble only turns positive during NBER recessions.

\begin{figure}[t]
\subcaptionbox{Selected recession classifier \label{f:classifier1978full}}{\smallgraphic[36]{\pdf}}\hfill
\subcaptionbox{Out-of-sample recession probability \label{f:oos1978full}}{\smallgraphic[37]{\pdf}}
\caption{Algorithm without retraining after 1978}
\note{Panel~A displays the classifier selected by the best method, R1, on April 1929--December 1977 data, over the entire sample, April 1929--August 2026: the purple line is the recession indicator, thin over the training period and thick after; the thin, pink lines are the thresholds. Panel~B displays the recession probability given by the classifier out of sample, January 1978--August 2026; the vertical, pink lines mark the NBER announcements of recession starts. The recession probability is given by \eqref{e:pkt}. Shaded areas indicate recessions dated by the \citet{NBER23}.}
\label{f:1978full}\end{figure}

\subsection{No retraining after 1995}

Kept unchanged, the F1 classifier selected in 1995 is the classifier used by the retrained algorithm until 2025 (figure~\ref{f:oos1995full}). Between 1995 and 2025, the classifier generates no false negatives but it produces a false positive. It detects the three recessions of 2001--2020 early: its early warnings come 1.3 months after the recession onset on average, and 5.7 months before the NBER announcement. Its final warnings come 4.0 months after the onset on average, and 4, 4, and 1 months before the NBER announcements. But it takes the 2023--2024 slowdown for a recession: its recession probability turns positive in October 2023, and reaches 100\% in July 2024. It turns positive again in September 2025, peaking at 12\%. 

The algorithm selects the same classifier at the start of the 1995--2004, 2005--2014, 2015--2021, and 2022--2025 tests (table~\ref{t:classifiers}). So the algorithm that stops retraining in 1995 produces exactly the same recession probability from 1995 to 2025 as the algorithm that retrains; and the algorithms that stop retraining in 2005, 2015, or 2022 produce segments of that same probability.

\begin{figure}[p]
\subcaptionbox{A classifier from the selected ensemble\label{f:classifier1985full}}{\smallgraphic[38]{\pdf}}\hfill
\subcaptionbox{Out-of-sample recession probability \label{f:oos1985full}}{\smallgraphic[39]{\pdf}}
\caption{Algorithm without retraining after 1985}
\note{Panel~A displays the first of the 5 classifiers selected by the best method, F5, on April 1929--December 1984 data (table~\ref{t:classifiers}), over the entire sample, April 1929--August 2026: the purple line is the recession indicator, thin over the training period and thick after; the thin, pink lines are the thresholds. Panel~B displays the recession probability given by the 5 classifiers out of sample, January 1985--August 2026; the vertical, pink lines mark the NBER announcements of recession starts. The thin, orange lines are the recession probabilities given by the individual classifiers, computed with \eqref{e:pkt}; the thick, purple line is their average, given by \eqref{e:pt}. Shaded areas indicate recessions dated by the \citet{NBER23}.}
\label{f:1985full}\end{figure}

\begin{figure}[p]
\subcaptionbox{Selected recession classifier  \label{f:classifier1995full}}{\smallgraphic[40]{\pdf}}\hfill
\subcaptionbox{Out-of-sample recession probability \label{f:oos1995full}}{\smallgraphic[41]{\pdf}}
\caption{Algorithm without retraining after 1995}
\note{Panel~A displays the classifier selected by the best method, F1, on April 1929--December 1994 data, over the entire sample, April 1929--August 2026: the purple line is the recession indicator, thin over the training period and thick after; the thin, pink lines are the thresholds. The algorithm selects the same classifier at the start of the 1995--2004, 2005--2014, 2015--2021, and 2022--2025 tests, so the figure also describes the algorithm without retraining after 2005, 2015, or 2022. Panel~B displays the recession probability given by the classifier out of sample, January 1995--August 2026; the vertical, pink lines mark the NBER announcements of recession starts. The recession probability is given by \eqref{e:pkt}. Shaded areas indicate recessions dated by the \citet{NBER23}.}
\label{f:1995full}\end{figure}

\end{document}